\documentclass[]{elsart}

\usepackage{icarus}

\usepackage{natbib}
\usepackage{longtable}

\bibpunct{(}{)}{;}{a}{,}{,}

\usepackage{graphicx}

\usepackage{amssymb}
\usepackage{wasysym}

\newcommand{\BibTeX}{ \textrm{B\kern-.05em\textsc{i\kern-.025em b}\kern-.08em
    T\kern-.1667em\lower.7ex\hbox{E}\kern-.125emX} }

\begin{document}

\begin{frontmatter}



\title{A Spectroscopic Comparison of HED Meteorites and V-type Asteroids in the Inner Main Belt}


\author[ifa,dtm]{Nicholas A. Moskovitz}, 
\author[ifa]{Mark Willman},
\author[bates]{Thomas H. Burbine},
\author[mit]{Richard P. Binzel},
\author[hilo]{Schelte J. Bus},

\address[ifa]{Institute for Astronomy, 2680 Woodlawn Drive, Honolulu, HI 96822 (U.S.A)}
\address[dtm]{Carnegie Institution of Washington, Department of Terrestrial Magnetism, 5241 Broad Branch Road, Washington, DC 20008 (U.S.A)}
\address[bates]{Departments of Geology and Physics \& Astronomy, Bates College, Lewiston, Maine 04240 (U.S.A)}
\address[mit]{Department of Earth, Atmospheric, and Planetary Sciences, Massachusetts Institute of Technology, Cambridge, MA 02139 (U.S.A)}
\address[hilo]{Institute for Astronomy, 640 North A'ohoku Place, Hilo, HI 96720 (U.S.A)}

\begin{center}
\scriptsize
Copyright \copyright\ 2010 Nicholas A. Moskovitz
\end{center}


%
%
%
%
%


\end{frontmatter}



\begin{flushleft}
\vspace{1cm}
Number of pages: \pageref{lastpage} \\
Number of tables: \ref{lasttable}\\
Number of figures: \ref{lastfig}\\
\end{flushleft}


\begin{pagetwo}{Comparison of V-type Asteroids and HED Meteorites}

Nicholas A. Moskovitz\\
Carnegie Institution for Science\\
Department of Terrestrial Magnetism\\
5241 Broad Branch Road\\
Washington, DC 20008, USA. \\
\\
Email: nmoskovitz@dtm.ciw.edu\\
Phone: (202) 478-8862 \\

\end{pagetwo}

\begin{abstract}

V-type asteroids in the inner Main Belt ($a<2.5$ AU) and the HED meteorites are thought to be genetically related to one another as collisional fragments from the surface of the large basaltic asteroid 4 Vesta.  We investigate this relationship by comparing the near-infrared ($0.7-2.5~\mu m$) spectra of 39 V-type asteroids to laboratory spectra of HED meteorites. The central wavelengths and areas spanned by the 1 and 2 $\mu m$ pyroxene-olivine absorption bands that are characteristic of planetary basalts are measured for both the asteroidal and meteoritic data. The band centers are shown to be well correlated, however the ratio of areas spanned by the 1 and 2 $\mu m$ absorption bands are much larger for the asteroids than for the meteorites. We argue that this offset in band area ratio is consistent with our currently limited understanding of the effects of space weathering, however we can not rule out the possibility that this offset is due to compositional differences. Several other possible causes of this offset are discussed. 

Amongst these inner Main Belt asteroids we do not find evidence for non-Vestoid mineralogies. Instead, these asteroids seem to represent a continuum of compositions, consistent with an origin from a single differentiated parent body.  In addition, our analysis shows that V-type asteroids with low inclinations ($i<6^\circ$) tend to have band centers slightly shifted towards long wavelengths. This may imply that more than one collision on Vesta's surface was responsible for producing the observed population of inner belt V-type asteroids. Finally, we offer several predictions that can be tested when the Dawn spacecraft enters into orbit around Vesta in the summer of 2011.

\end{abstract}



\section{Introduction \label{sec.intro}}

The basaltic howardite, eucrite and diogenite meteorites (HEDs) and the large Main Belt asteroid 4 Vesta have traditionally been linked due to their spectroscopic similarity and the lack of any other large asteroid with the characteristic spectral signature of magmatic basalts \citep[][]{1970Sci...168.1445M,1977GeCoA..41.1271C}. The presence of a collisional family associated with Vesta supports this link. Vesta-family members, often referred to as the Vestoids, are spectroscopically classified as V-types and are dynamically linked to Vesta. The term non-Vestoid refers to any V-type asteroid that originated on a parent body other than Vesta.

\begin{figure}[]
\begin{center}
\includegraphics[width=14cm]{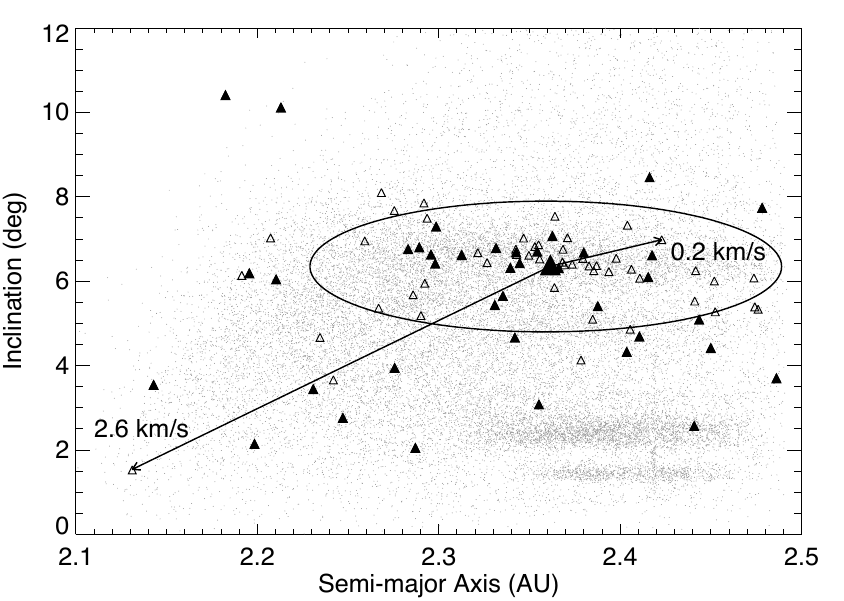}
\end{center}
\caption[Dynamical map of V-type asteroids in the inner Main Belt]{
Dynamical map in proper element space of V-type asteroids in the inner Main Belt. The filled triangles represent the V-type asteroids with NIR spectra included in this study. The open triangles are all spectroscopically confirmed V-types based solely on visible wavelength data \citep{1995Icar..115....1X,2002Icar..158..146B,2004Icar..172..179L,2006A&A...459..969A,2008Icar..198...77M}. Vesta is denoted by the large filled triangle at 2.36 AU, 6.35$^\circ$. The grey dots in the background are all objects from the 4th release of the SDDS MOC \citep{2008Icar..198..138P}. The ellipse traces the 0.6 km/s ejection velocity relative to Vesta. The relative ejection velocities for two objects are indicated by the arrows.
} 
\label{fig.inner}
\end{figure}

The Vestoids extend from the $\nu_6$ secular resonance at the inner edge of the Main Belt to the 3:1 mean motion resonance with Jupiter at 2.5 AU \citep[Fig. \ref{fig.inner},][]{1993Sci...260..186B}. These resonances act as a dynamical escape hatch from the Main Belt and can transport fragments removed from the surface of Vesta (or one of the Vestoids) to the Earth as HED meteorites \citep{1997Sci...277..197G}. Resolved images of Vesta reveal a large crater ($\sim460$ km in diameter) on its south pole  \citep{1997Sci...277.1492T}, supporting a scenario of collisional formation for the Vesta family. 

Hydrocode simulations of the collision that formed the Vesta family \citep{1997M&PS...32..965A} suggest that km-size fragments would have been removed with ejection velocities ($\Delta v$) of no greater than approximately 0.6 km/s. Simplified versions of Gauss's equations can be used to quantify the distance from Vesta in orbital element space corresponding to this ejection velocity \citep{1996Icar..124..156Z}.\footnote{The following calculations assume that the true anomaly and argument of perihelion at the time of formation of the Vesta family were equal to those calculated by \citet{1996Icar..124..156Z}. However, these authors did not consider the effects of orbital migration due to the Yarkovsky force \citep{2006AREPS..34..157B}. Therefore, the calculations presented here are approximations accurate to $\sim20\%$.} The maximum range of semi-major axes for collisionally-produced fragments can be estimated by assuming that a hypothetical Vestoid was ejected with a 0.6 km/s velocity vector aligned exclusively in a direction tangential to its orbit  \citep[Equation 4 in][]{1996Icar..124..156Z}. This calculated range is $2.23-2.49$ AU, centered on the semi-major axis of Vesta (2.36 AU). Similar calculations can be done for both eccentricity and inclination, producing ranges of $0.06-0.13$ and $4.8-7.9^\circ$ respectively. This region of orbital element space is enclosed by the ellipse in Figure \ref{fig.inner}.

Figure \ref{fig.inner} shows numerous V-type asteroids in the inner Main Belt with values of $\Delta v$ much larger than 0.6 km/s, some with values in excess of 2 km/s. When these objects were first discovered \citep[e.g.][]{1993Sci...260..186B,2001M&PS...36..761B,2002Icar..159..178F,2004Icar..172..179L,2006A&A...459..969A} it was unclear how they could have reached such orbits. However, recent progress in the use of numerical integrators has helped to clarify this issue. \citet{2005A&A...441..819C} showed that three-body and weak secular resonances could lead to the migration of some Vestoids to orbits with $\Delta v>0.6$ km/s. \citet{2008Icar..193...85N} showed that a combination of these resonances and the Yarkovsky effect could disperse the orbits of Vestoids to nearly the full extent of the inner Main Belt. However, these authors found that the observed number of V-type asteroids at low inclination ($i<6^\circ$) was too large to be explained by their model of Vestoid migration. 

Three possibilities exist to explain this over-abundance of low-$i$ V-types. First, these objects could be fragments of basaltic crust from a non-Vestoid  differentiated parent body. In this case these objects could be spectroscopically distinct from the Vestoids, as is the case for non-Vestoid V-types in the outer Main Belt \citep[e.g. ][]{2000Sci...288.2033L,2008ApJ...682L..57M}. Second, they may be from Vesta, but were removed from the surface before the Late Heavy Bombardment (LHB), before the primary family forming collision, and were scattered to their current orbits as mean motion and secular resonances swept through the Main Belt during the LHB \citep{1997AJ....114..396G}. In this case these objects would represent an older population of Vestoids, removed from a different region on Vesta's surface and thus might be spectroscopically distinct. Third, these objects may have been ejected from the Vesta parent body at the time of family formation and have since migrated to their current orbits by some unexplored dynamical mechanism. In this case these objects should not appear spectroscopically different from other V-type asteroids in the inner Main Belt. 

A similar line of reasoning motivated \citet{1998AMR....11..163H} to investigate the visible-wavelength spectral features (namely the slope and 1 $\mu m$ band depth) of 20 V-type asteroids in the inner Main Belt as a function of their orbital elements. These authors found that V-type asteroids with large values of $\Delta v$ tended to have steeper spectral slopes than V-types with smaller $\Delta v$. However, the largest value of $\Delta v$ considered by these authors was 0.65 km/s, very close to the expected ejection velocity of Vestoid fragments. Furthermore, all of the objects that were studied have since been incorporated into the Vesta dynamical family as detection completeness has increased in the last decade. Thus it is surprising that this spectroscopic trend was observed as a function of orbital parameters for objects that plausibly originated at the same time from the same parent body. 

\citet{2004Icar..171..120D} attempted a similar investigation into the spectroscopic diversity of Vestoids at NIR wavelengths. This study produced unexpected results: the Band II to Band I area ratios and the Band I and Band II centers for the majority of the V-types in their sample did not agree with those of the HEDs (Band I and II refer to the 1 and 2 $\mu m$ absorption bands common to basaltic material, see \S\ref{sec.tools} for definitions of these parameters). Although it has been suggested that band area ratios are sensitive to variations in grain size, temperature and space weathering \citep{2002LPI....33.2023U}, band centers should be less sensitive to these effects and thus comparable between genetically related populations (i.e. the HEDs and Vestoids).

In light of recent advances in dynamical simulations \citep{2005A&A...441..819C,2008Icar..193...85N} and improvements in NIR spectroscopic instrumentation \citep{2003PASP..115..362R}, we revisit the issue of the diversity of basaltic asteroids by measuring the NIR spectral properties of 39 inner Main Belt V-type asteroids. The goals of this study are threefold: (1) address the reported spectro-dynamical correlation amongst V-type asteroids in the inner Main Belt \citep{1998AMR....11..163H} by extending our analysis out to NIR wavelengths and by including V-types across a wider range of orbital element space; (2) address the findings of \citet{2004Icar..171..120D} to look for spectroscopic differences between inner belt V-type asteroids and HED meteorites; (3) determine if any of the V-type asteroids in the inner Main Belt have spectroscopic properties suggestive of a non-Vestoid mineralogy. It is important to note that we do not attempt to extract detailed mineralogical information for individual asteroids. Instead we characterize our relatively large data set using band analysis techniques \citep[e.g.][]{1986JGR....9111641C} with the intent of making statistically significant statements about the gross spectral properties of V-type asteroids relative to those of the HED meteorites.

%
%
\section{Spectral Band Analysis \label{sec.tools}}

\begin{figure}[]
\begin{center}
\includegraphics[width=14cm]{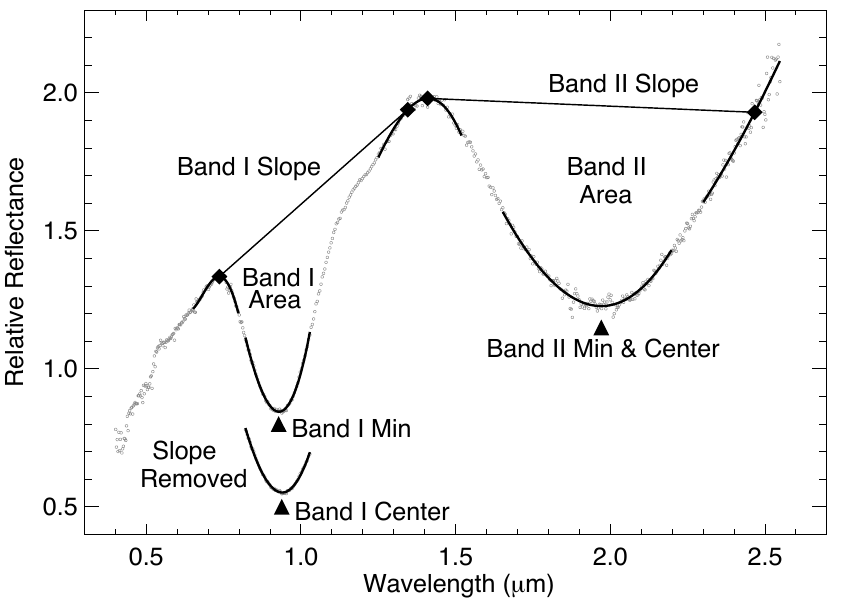}
\end{center}
\caption[Band analysis of V-type asteroid 5498 Gustafsson]{
Spectrum and band parameters of V-type asteroid 5498 Gustafsson. The data (grey circles) have been normalized to unity at 1.0 $\mu m$. The slope-removed data plot below the normalized spectrum. The band parameters of interest are denoted: Band I and II minima, centers, areas and slopes. This figure is based upon \citet{1986JGR....9111641C}. The filled diamonds mark the edges and tangent points for the Band I and II slopes. Note that all band parameters are defined relative to the fitted segments (solid black lines), not the data.
}
\label{fig.bands}
\end{figure}

The reflectance spectra of minerals contain absorption features that are diagnostic of properties such as composition, albedo, grain size and crystal structure. A variety of analytic tools have been developed to interpret such information from remotely obtained spectra \citep[e.g.][]{1990JGR....95.6955S,1993Icar..102..107H,1999Icar..137..235S,2007JGRE..11207005L}. Band analysis techniques \citep[e.g.][]{1986JGR....9111641C} are useful for characterizing spectral data and can be used as a starting point for deriving mineralogical information \citep{2002aste.conf..183G}. As a characterization tool, band analyses are completely objective, requiring no knowledge or assumptions about mineralogy, and can be implemented quickly for a large number of spectra. The primary goal of this work is to use band analysis techniques to facilitate a comparison between HEDs and V-type asteroids. We perform this band analysis with attention to the following parameters: the minima and central wavelength of the 1 and 2 $\mu m$ olivine-pyroxene absorption features (Band I and II respectively) and the ratio of areas within these two bands. These key parameters are depicted in Figure \ref{fig.bands}. We also measure the Band I slopes of the asteroids for comparison to results of \citet{1998AMR....11..163H}.

Measurement of these parameters is dependent on several polynomial fits to actual data (Fig. \ref{fig.bands}). These fits include a fourth order fit to the reflectance peak at $\sim0.7~\mu m$, third order fits to the reflectance troughs and peak at $\sim0.9,~2.0$ and $1.4~\mu m$, and a second order fit to the red-edge of the $2~\mu m$ band. The band minima are equal to the minima of the third order fits across each of the reflectance troughs ($0.82-1.03~\mu m$ for Band I and $1.65-2.2~\mu m$ for Band II). The Band I center is measured by first fitting a line (labelled Band I Slope in Fig. \ref{fig.bands}) that intersects the fitted peak at $\sim0.7~\mu m$ and is tangent to the reflectance peak at $\sim1.4~\mu m$. This continuum slope is then divided into the data and the minimum of this slope-removed band is defined as the band center. Band I centers are typically at longer wavelengths than Band I minima because of the positive Band I slope for most HEDs and V-type asteroids. Band II centers are defined as equal to the Band II minima (without dividing out a continuum slope), because the red edge of the $2~\mu m$ band is typically unresolved in NIR telescopic data (e.g.~Fig. \ref{fig.bands}). We define the red edge of the $2~\mu m$ band to be at $2.44~\mu m$ (the reddest wavelength for which all of our data are reliable). Band areas are calculated as the area between the band slopes and data points. The calculation of error bars for each of these band parameters is discussed in \S\ref{sec.vtype_bands}.

We have intentionally been very explicit in our definition of these band parameters because other studies have employed different methodologies. For instance, \citet{1986JGR....9111641C} define the red edge of Band II at 2.4 $\mu m$ while \citet{2002aste.conf..183G} define it at 2.5 $\mu m$. In some studies, the band centers are defined as the wavelength at which the band areas are bisected \citep[e.g.][]{2009Icar..200..480B}. Although these differences are subtle, they do make it difficult to compare measurements from one study to another \citep[a similar discussion of this issue was also presented by][]{2006AdSpR..38.1987D}. We apply our methodology to all of the data considered here so that comparisons can be made between the asteroids and meteorites. These measurements are internally consistent and are not meant to be directly compared to previous works.

%
%
\section{Band Analysis of HED Meteorites}

To build a comparison database for the asteroids we perform a spectral band analysis for each of the HED meteorites that are included in the Brown RELAB spectral catalog \citep{2004LPI....35.1720P}. All samples that were analyzed are listed in the Appendix. We safely ignore error bars in the measurement of the meteorite band parameters, because they are much smaller than those associated with the asteroidal data (\S\ref{sec.vtype_bands}), and they become irrelevant in our statistical approach to characterizing the overall spectral properties of the HEDs. The range of HED band parameters roughly agree (within $\sim10\%$) with those measured by previous investigators \citep[][]{2005M&PS...40..445D}.

This set of HED spectra excludes some of the data that are available in the RELAB database. We do not include samples that have been processed in laser irradiation experiments. When multiple spectra for a single sample exist, which is true for about 40\% of the samples, we choose data corresponding to the smallest available grain size (for RELAB this is $25\mu m$). This approach is reasonable because the grains that compose the regoliths of Vestoids are believed to be $<25\mu m$ in size \citep{1994Metic..29..394H,1995Icar..115..374H}. We note that preferentially selecting these small-grained samples does not have a significant effect on any of our results. When relevant (e.g. \S\ref{sec.BARs}), we discuss the implications of this selection criterion. Our aim in analyzing these meteoritic spectra is not to perform a detailed study of spectral changes as a function of particle size, irradiation level and other variables; for such a study see \citet{2005M&PS...40..445D}.

\begin{figure}[]
\begin{center}
\includegraphics[width=14cm]{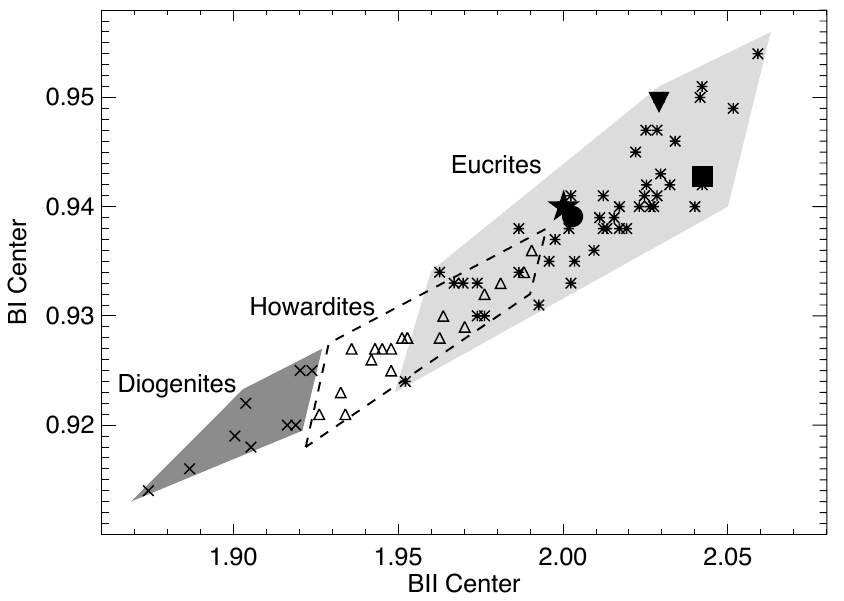}
\end{center}
\caption[Band centers of HED meteorites]{
Band centers of the howardites (triangles), eucrites (asterisks) and diogenites ($\times$'s). Approximate boundaries for each of these subgroups are depicted as the dashed line, light grey and dark grey regions respectively. The eucrites and diogenites are very clearly segregated in this figure. The parameters of the four isotopically anomalous eucrites that are included in our study are denoted by the filled symbols: star (Ibitira), circle (Passamonte), upside-down triangle (NWA011), and square (PCA91007).} 
\label{fig.B1B2_HED}
\end{figure}

A plot of BI center versus BII center for the HEDs is shown in Figure \ref{fig.B1B2_HED}. The three HED subgroups are well defined (although they do overlap) in this parameter space. The regions occupied by each subgroup have been shaded or outlined so that they can be compared to the band parameters of the V-type asteroids in \S\ref{sec.vtype_bands}. In general, eucrites tend to have band centers at longer wavelengths than the diogenites, which implies that they are more Fe-rich and have Ca/Mg abundance ratios much greater than 1 \citep{2001M&PS...36..761B}. This is confirmed by chemical analyses of eucrite samples \citep{1998psc..book.....L}. The diogenites have shorter wavelength band centers due to Ca/Mg abundance ratios much less than 1 and relatively Fe-poor compositions. Unsurprisingly, the howardites (brecciated mixes of eucrites and diogenites) have band centers that are intermediate to that of the eucrites and diogenites. The compositional trend represented by the band centers is consistent with the origin of the diogenites at greater depths within the HED parent body, i.e. low-Ca diogenites crystallized and settled before high-Ca eucrites as the HED parent body cooled \citep{1997M&PS...32..825R}.

\begin{figure}[]
\begin{center}
\includegraphics[width=14cm]{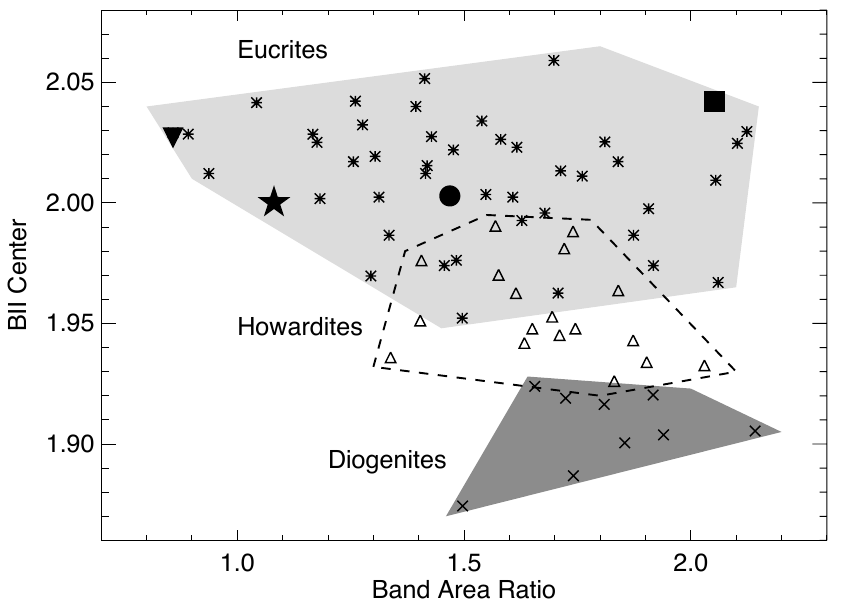}
\end{center}
\caption[Band area ratios of HED meteorites]{
BAR versus BII center for HED meteorites. The symbols and regions are the same as defined in Figure \ref{fig.B1B2_HED}. The three HED subgroups are not clearly segregated in BAR, though diogenites do tend towards larger values.} 
\label{fig.BAR_HED}
\end{figure}

A plot of BAR versus BII center is shown in Figure \ref{fig.BAR_HED}. Traditionally, BAR is plotted versus BI center due to the mineralogical relevance of this particular combination of axes \citep{1993Icar..106..573G}. However, we are not attempting to draw mineralogical inferences from our data. Furthermore, we find that the BII centers of the three HED subgroups are better segregated with less overlap than the BI centers. Thus, the axes in Figure \ref{fig.BAR_HED} are the preferred method for presenting our data. 

The BARs do not clearly segregate the three HED subgroups, however the diogenites do tend towards larger values. The lack of high-Ca pyroxene amongst the diogenites \citep{2001M&PS...36..761B} is a likely cause for their higher BARs \citep{1986JGR....9111641C}. Truncating the edge of BII at 2.44 $\mu m$ might also have affected their BARs.

Of the 75 HEDs that were analyzed, four eucrites (Passamonte, Ibitira, NWA011, and PCA91007)  have anomalous oxygen isotope ratios. Specifically, these meteorites all have [$^{17}$O/$^{16}$O] abundance ratios that are at least $0.03\permil$ away from the mean [$^{17}$O/$^{16}$O] ratios for ordinary HEDs, which display a scatter of only $\pm~0.016 \permil$ relative to their mean \citep{2009LPI....40.2263S}. The interpretation for these four eucrites is that they derive from unique (non-Vestoid) differentiated parent bodies. Our analysis shows that the band paramters of these four samples are in no way different from the other eucrites (Figs.~\ref{fig.B1B2_HED} \& \ref{fig.BAR_HED}). At face value this would seem to suggest that the spectroscopic properties (and by extension the mineralogy) of basaltic material from one parent body to the next is indistinguishable and that isotopic information is needed to distinguish individual parent bodies. However, this does not seem to be the case for all V-type asteroids in the Main Belt. For example, 1459 Magnya is a non-Vestoid V-type with distinct band parameters \citep[][]{2004Icar..167..170H}. Thus, while band analysis techniques are capable of distinguishing V-type asteroids with no genetic relation to Vesta, this does not seem to be the case amongst the HEDs. The reason for this remains unclear.

%
%
\section{Band Analysis of V-type Asteroids \label{sec.vtype_bands}}

\subsection{Observation \& Reduction}

We collected NIR spectra of 39 V-type asteroids distributed throughout the inner Main Belt (Fig.~\ref{fig.inner}, Appendix A, Table \ref{tab.obs}). All of these observations, with the exception of the data of Vesta from 1981, were obtained with the SpeX instrument \citep{2003PASP..115..362R} at NASA's IRTF. In addition to our new observations, these data include those  from \citet{2009Icar..202..160D}, data from the ongoing MIT-UH-IRTF Joint Campaign for NEO Spectral Reconnaissance (NEOSR, PI's Richard Binzel \& Alan Tokunaga) and the spectrum of 4 Vesta from \citet{1997Icar..127..130G}. See the specific references for details regarding observing conditions and reduction of these data sets.

All new SpeX observations were made with the telescope operating in a standard ABBA nod pattern. SpeX was configured in its low resolution (R=250) prism mode with a 0.8" slit for wavelength coverage from $0.7 - 2.5~\mu m$.  All targets were observed at an air mass of less than 1.5 and as near as possible to their meridian crossings. The slit was orientated along the parallactic angle at the start of each observation to minimize the effects of atmospheric dispersion. Exposure times were limited to 200 seconds. The number of exposures was selected in an attempt to obtain a consistent $S/N\sim100$ for all asteroids. Internal flat fields and arc line spectra were obtained immediately after or before each target asteroid. A solar analog star was observed to calibrate each asteroid spectrum (i.e. solar correction, compensating for instrument response and removal of atmospheric absorption features). Analogs were observed closely in time, altitude and azimuth to the asteroid. The spectrum of each asteroid was calibrated using a single solar analog. Specific analogs were selected as those which minimized residual atmospheric features in the calibrated spectrum of a given asteroid. Calibration and reduction of our new observations employed the Spextool package \citep{2004PASP..116..362C}.

When available, visible wavelength data from \citet{2008Icar..198...77M}, SMASS \citep{1995Icar..115....1X,2002Icar..158..106B} or S3OS2 \citep{2004Icar..172..179L} were appended to the NIR spectra to increase wavelength coverage and to improve S/N at the blue edge of the 1 $\mu m$ absorption band. The addition of visible wavelength information to the NIR spectra does not significantly effect the trends in band parameters that are discussed in the following sections. Figure \ref{fig.obs} shows that both the 1 and 2 $\mu m$ absorption bands are resolved for all 49 of the spectra and that the typical S/N is either better than or comparable to previous studies of NIR band parameters of V-type asteroids \citep[e.g.][]{2003Icar..165..215K,2004Icar..171..120D,2004Icar..167..170H}. The targets 2653 Principia, 2851 Harbin, 3155 Lee, 3782 Celle and 4215 Kamo were each observed at two different epochs by the NEOSR program (Table \ref{tab.obs}). The data from these pairs of observations were combined into composite spectra to improve S/N and are presented as such in Figure \ref{fig.obs}. New observations of asteroids 2653 Principia and 4215 Kamo are also presented in Figure \ref{fig.obs}.

\subsection{Temperature Correction to Band Centers \label{sec.tempcorr}}

The band analysis procedure described in \S\ref{sec.tools} was applied to each of the NIR asteroidal spectra. The results of this band analysis are summarized in Table \ref{tab.band_params}. In order to compare these parameters to those of the HEDs, which were measured in a laboratory at $\sim300$ K, a temperature correction to the band centers was required. Temperature corrections were computed following the methodology of \citet{2009M&PS...44.1331B}. The mean surface temperature of an asteroid ($T$) is approximated by the equation for energy conservation:
\begin{equation}
T = \displaystyle\left[\frac{(1-A)L_0}{16\eta\epsilon\sigma\pi r^2}\right]^{1/4},
\label{eqn.temp}
\end{equation}
where $A$ is the albedo, $L_0$ is the solar luminosity ($3.827\times10^{26}$ W), $\eta$ is the thermal beaming parameter (assumed to be unity), $\epsilon$ is the asteroid's infrared emissivity \citep[assumed to be 0.9,][]{2005Icar..173..385L}, $\sigma$ is the Stefan-Boltzman constant and $r$ is the asteroid's heliocentric distance. Though the albedo of 4 Vesta and 1459 Magnya are approximately 0.4 \citep{2004PDSS...12.....T,2006Icar..181..618D}, recent Spitzer observations suggest that smaller V-type asteroids may have albedos as low as 0.2 \citep{2009LPI....40.2204L}. Thus, the albedo of each target was assumed to be 0.3. The calculated mean surface temperature for each of the V-type asteroids is given in Table \ref{tab.band_params}. Uncertainties in the adopted parameters ($A$, $\eta$ and $\epsilon$) do not have a large affect on our resulting temperature corrections because of the small (1/4) exponent in Equation \ref{eqn.temp}.

The temperature-dependent wavelength correction (in microns) to the BI and BII centers are given by \citep{2009M&PS...44.1331B}:
\begin{equation}
\Delta BI = (5.05\times10^{-3}) - (1.70\times10^{-5}) T,
\label{eqn.db1}
\end{equation}
\begin{equation}
\Delta BII = (5.44\times10^{-2}) - (1.85\times10^{-4}) T.
\label{eqn.db2}
\end{equation}
These equations are derived from linear fits to laboratory measured band centers for two different pyroxenes at temperatures of 293, 173 and 80 K \citep{2000Icar..147...79M}. In general these corrections are small: 0.002 $\mu m$ for all of the $\Delta BI$ values (rounded to the resolution limit of the spectra) and about an order of magnitude more for $\Delta BII$ (Table \ref{tab.band_params}). These shifts are much less than the range of measured band centers and thus do not have a major impact on our results, however they do improve the measured overlap of the HEDs and V-types (see \S\ref{sec.diagrams}). The temperature corrected band centers are presented in Table \ref{tab.band_params}.

\subsection{Estimated Error Bars \label{sec.errorbars}}

The spectrum of asteroid 36412 (2000 OP49) has the lowest wavelength-averaged signal-to-noise ratio (S/N$\sim$27) in this data set. This object's spectrum was used to estimate maximum statistical error bars for the measured asteroidal band parameters. Determining the error bars on band centers is non-trivial because the uncertainties for the observed data points are along the reflectance axis, not along the wavelength axis. To access the errors on the band centers of asteroid 2000 OP49 we ran a Monte Carlo simulation in which random noise (at the level of the measured signal-to-noise) was added to segments of the spectrum in the vicinity of the Band I and Band II minima. These segments were re-fit with a third-order polynomial and the new band centers were calculated. This process was repeated 100,000 times and the standard deviation of the measured band centers was calculated. The result of this error analysis gives $\sigma_{BI}=0.003~\mu m$ and $\sigma_{BII}=0.011~\mu m$. This procedure for estimating errors on band centers is similar to that of \citet{2007DPS....39.2003S}. Most of the asteroidal spectra are of much higher S/N ($\sim100$) and thus will have statistical errors for their band centers that are insignificant relative to the general trends in which we are interested.

The errors for BARs are calculated by first measuring the largest and smallest possible band areas based on the S/N characteristics of a given spectrum. The maximum band areas are measured by subtracting noise at the 1-sigma level from each data point, while the  minimum band areas are measured by adding noise at the 1-sigma level. For example, the BII area of 2000 OP49 with 1-sigma worth of noise subtracted from the data points is 0.608 and the BII area with 1-sigma worth of noise added is 0.516. We define the errors on the individual band areas to be one half of the difference between these maximum and minimum values. The net error on the BAR then involves the propagation of the errors from each band. Performing this analysis on the spectrum of 2000 OP49 gives a maximum error bar on the asteroidal BARs of $\sigma_{BAR}=0.23$. 

\subsection{Band Diagrams \label{sec.diagrams}}

The asteroidal band centers are plotted in Figure \ref{fig.B1B2_Vtypes} along with the howardite, eucrite and diogenite regions that were defined for Figure \ref{fig.B1B2_HED}. Eight asteroids (4 Vesta, 956 Elisa, 1468 Zomba, 2045 Peking, 2579 Spartacus, 2653 Principia, 2763 Jeans and 4215 Kamo) were observed more than once. These individual observations are connected by the thin lines. The maximum statistical error bars calculated from the spectrum of 2000 OP49 (\S\ref{sec.errorbars}) are indicated in this figure and are clearly much smaller than the size of the HED regions and the range of band centers displayed by the asteroids. The variability between multiple observations of individual asteroids is typically larger than the statistical errors and is likely due to a combination of compositional variation across the surface of the asteroids, differences in observing circumstances (e.g. use of different solar analogs, different observational phase angles), and uncorrected systematic effects. Vesta is known to display large variation in its band centers due to compositional variation on its surface \citep{1997Icar..127..130G}. However, such variation may not apply to these targets, which are more than an order of magnitude smaller in size. Observations of single asteroids through multiple rotation periods and with multiple instruments could determine whether any systematic effects may be inherent to these SpeX data. Note that the mean of the band centers for the V-types do not seem to be affected by the (apparently) random variability introduced by multiple observations of individual objects.

Four of the asteroids [1468 Zomba, 2823 van der Laan, 3703 Volkonskaya and 50098 (2000 AG98)] were observed on single epochs as part of the NEOSR program and do not have complementary visible wavelength data. These data have been truncated at 0.82 $\mu m$ as part of the standard NEOSR reduction protocol. Thus their full 1 $\mu m$ absorption feature is not resolved, thereby preventing a calculation of a BI slope and center. However, an upper limit to the BI center can be made based on the measured BI minima. The largest offset between a BI minimum and a BI center was measured to be 0.016 $\mu m$ for asteroid 7800 Zhongkeyuan. Adding this value to the BI minima of the four NEOSR asteroids produces reasonable upper limits to their BI centers. These four objects are plotted in Figure \ref{fig.B1B2_Vtypes} as upside-down, filled triangles.

As previous authors have noted \citep{2001M&PS...36..761B}, there appears to be a lack of diogenetic V-type asteroids in the inner Main Belt. However, we do find four objects that plot near the howardite-diogenite boundary in Figure \ref{fig.B1B2_Vtypes}: 2851 Harbin, 3155 Lee, 26886 (1994 TJ2) and 27343 (2000 CT102). These objects and the lack of diogentie-like asteroids are discussed in greater detail in the following section.

Our measured band centers do not span the full range of this parameter space as found by \citet{2004Icar..171..120D}, instead our data suggests a good correlation between the band centers of the HEDs and V-type asteroids. This discrepancy may be due to the lower S/N of the \citet{2004Icar..171..120D} dataset and/or differences in band analysis techniques. 

The dominant effect of the applied temperature corrections (\S\ref{sec.tempcorr}) is a shift of the ensemble of asteroidal B2 centers by approximately +0.02 $\mu m$. This shift, though small, improves the overlap between the V-type points and HED regions in Figure \ref{fig.B1B2_Vtypes}.

A plot of asteroidal BAR versus BII center is shown in Figure \ref{fig.BAR_Vtypes}. Only objects with fully resolved 1 and 2 $\mu m$ bands can be plotted in this figure [NEOSR targets 1468 Zomba, 2823 van der Laan, 3703 Volkonskaya and 50098 (2000 AG98) are not included]. The plotted V-type asteroids have band centers that are offset from those of the HEDs by an amount that is much greater than the maximum statistical error bar. This discrepancy has been noted before \citep{2005M&PS...40..445D} and was attributed to several possible causes, including grain size, temperature, weathering and mineralogical effects. In \S\ref{sec.BARs} we present specific arguments for possible causes of this spectral mismatch.

%
%
\section{Spectroscopic Diversity of Inner Belt V-type Asteroids \label{sec.diversity}}

\begin{figure}[]
\begin{center}
\includegraphics[width=14cm]{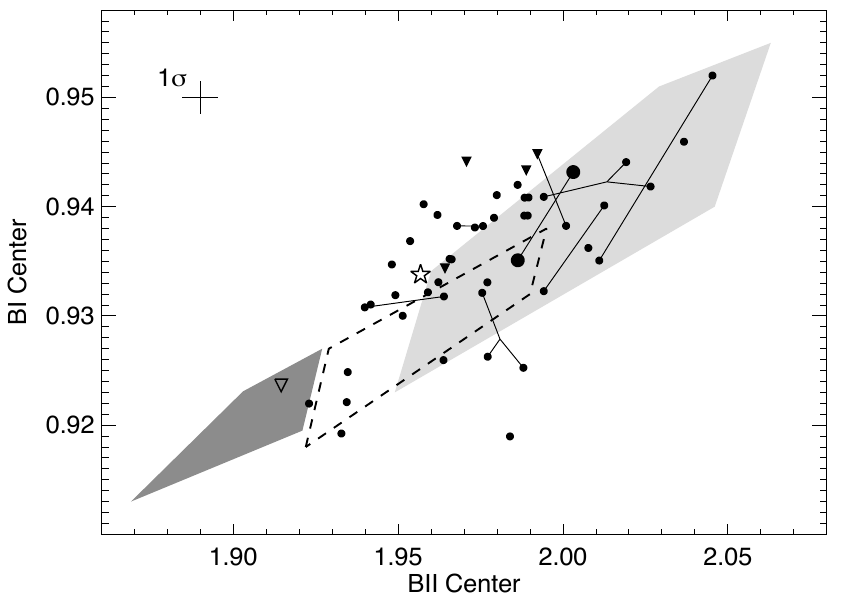}
\end{center}
\caption[Band centers of V-type asteroids]{
Measured band centers of V-type asteroids (solid circles). Maximal 1-sigma error bars calculated from the analysis of the spectrum of 2000 OP49 (\S\ref{sec.errorbars}) are shown in the upper left corner. The approximate boundaries for each of the HED subgroups are the same as in Figure \ref{fig.B1B2_HED}. Multiple observations of individual objects are connected by the thin lines. Vesta is plotted as the large filled circles. Upper limits to the BI centers for asteroids Zomba, Modena, Volkonskaya and 2000 AG98 are plotted as the upside-down filled triangles. The temperature-corrected band centers of non-Vestoid V-type Magnya (BI center = 0.934, BII center = 1.957) is plotted as a star. The Band II center and upper limit to the Band I center for non Vestoid V-type 1995 WV7 is plotted as an open upside-down triangle (lower left).
} 
\label{fig.B1B2_Vtypes}
\end{figure}

We begin this discussion by considering whether any of the inner belt V-type asteroids represent distinctly non-Vestoid mineralogies. This is facilitated by first considering the spectroscopic properties of two non-Vestoid V-type asteroids that orbit outside of the 3:1 mean motion resonance with Jupiter: 1459 Magnya and 21238 (1995 WV7). These are the only two non-Vestoid V-types with published NIR spectra. From the all-night average spectrum of \citet{2004Icar..167..170H} we compute the band parameters of Magnya: BI center = 0.933, BII center = 1.957, and BAR = 3.395. These band centers are plotted in Figure \ref{fig.B1B2_Vtypes} and are undistinguishable from the population of inner belt V-types that we have analyzed. However, the BAR of Magnya is much larger than any of the objects (asteroid or meteorite) that we have measured. This large BAR is similar to the value obtained by \citet{2004Icar..167..170H}.

The second of the non-Vestoid V-types is asteroid 21238 (1995 WV7). Unfortunately a full ($0.4-2.5~\mu m$) spectrum of this asteroid does not exist in published literature. The best available data ($0.83-2.46~\mu m$) for 21238 is from the NEOSR online database (http://smass.mit.edu/minus.html). From this truncated spectrum only band minima can be measured: BI minimum = 0.906, BII minimum = 1.889. Again, an upper limit to the BI center of 21238 can be calculated by assuming the maximum offset between its BI minimum and BI center (0.016). This enables the inclusion of 21238 in Figure \ref{fig.B1B2_Vtypes}, where it falls in the lower-left corner, somewhat isolated relative to the other V-type asteroids and the only object that directly overlaps the region occupied by the diogenites.

The spectroscopic properties of these two non-Vestoid V-types (BAR for 1459 Magnya and inferred band centers for 21238) are offset from the asteroids included in our sample, thus giving credence to the dynamical arguments that these must be non-Vestoid V-types \citep{2000Sci...288.2033L,2002Icar..158..343M,2007A&A...473..967C,2008Icar..194..125R}. Amongst our sample of inner Belt V-types, asteroid 4038 Kristina is significantly offset with a lower BI center compared to the general HED trend (though its BII center and BAR are in no way unusual). Though this unusual BI center might imply a non-Vestoid mineralogy, that would be surprising for Kristina because it resides closer to Vesta in orbital element space ($\Delta v=0.03$ km/s) than any other of our targets. For this reason we interpret its BI center as an indication of the mineralogical diversity inherent to members of the Vesta family and/or variability intrinsic to the observational circumstances for 4038 Kristina.

From a statistical standpoint, we find that the band centers of the HEDs and V-types are virtually indistinguishable: the mean B1 centers are 0.934 for both populations and the mean B2 centers are 1.976 and 1.987 respectively. Figure \ref{fig.B1B2_Vtypes} confirms that the band centers of these populations are in close agreement. This is consistent with their origin on a single parent body. 

It is unclear why a group of asteroids are shifted towards Band I centers that are slightly higher than those of the HEDs (Fig. \ref{fig.B1B2_Vtypes}). Amongst the HEDs, high Ca- and Fe-abundances push band centers to longer wavelengths \citep{2001M&PS...36..761B}, however this should affect both Band I and Band II. This shift could be explained if we have underestimated the temperature corrections to the band centers (Equations \ref{eqn.db1} and \ref{eqn.db2}). Another possibility may stem from the absence of visible wavelength data for some of these asteroids. This could be affecting the band analysis and pushing some objects towards longer BI centers. A final possibility is that this shift is a result of actual mineralogical differences, with V-type asteroids representing a wider range of compositions than are observed amongst the HEDs. 

Four objects [2851 Harbin, 3155 Lee, 26886 (1994 TJ2) and 27343 (2000 CT102)] have higher than average BARs and are offset towards smaller band centers from the majority of the asteroids in Figure \ref{fig.B1B2_Vtypes}. This apparent offset in band center is likely due to incomplete number statistics rather than the presence of non-Vesta mineralogies. Future observations of additional Vestoids would likely fill in the band center gap between these four objects and the rest of the objects studied here.

The lack of diogenite-like bodies in our sample of 39 asteroids is puzzling. This has been previously noted for V-types in the Main Belt \citep{2001M&PS...36..761B} and in the NEO population \citep{2009M&PS...44.1331B}. It is plausible that the four asteroids near the diogenite border in Figure \ref{fig.B1B2_Vtypes} are primarily composed of diogenites with some eucritic material contaminating their surfaces. In general, the Vestoids are probably rubble-pile aggregates of collisional debris.  The removal of large fragments from diogenetic depths ($\sim10$s of km) within the Vesta parent body may not have occurred. Thus, fragments of diogenite-like material on the surfaces of V-type asteroids may only be observable with high resolution surface maps and will remain undetected in the hemispherical averages provided by telescopic data. An argument that is consistent with a rubble pile model for the Vestoids is that the distribution of cosmic ray exposure ages is similar for each sub-group of the HED meteorites \citep{2006mess.book..829E}.

Lastly, we note that the asteroid 2579 Spartacus was found to have a smaller BAR ($\sim$1.3) than the rest of our sample (Fig. \ref{fig.BAR_Vtypes}). This has been noted before \citep{2001M&PS...36..761B} and is thought to be related to an anomalously high olivine content relative to other Vestoids. This could be due to an origin from deep within the Vesta parent body where olivine abundances were higher than in the eucritic crust, or it could be due to its origin on another parent body. If Spartacus was removed from large depths within the Vesta parent body then we would expect other diogenetic asteroids to also be present in the Main Belt, perhaps at sizes below our current observational limits.

%
%
\section{Asteroidal Versus Meteoritic BARs \label{sec.BARs}}

\begin{figure}[]
\begin{center}
\includegraphics[width=14cm]{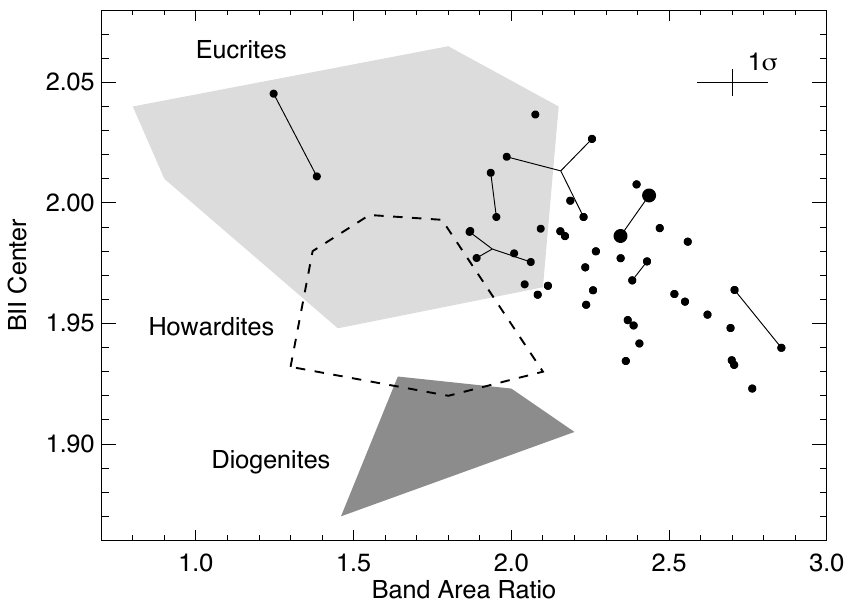}
\end{center}
\caption[Band area ratios of V-type asteroids]{
BAR versus BII center for V-type asteroids. The maximum 1-sigma error bars that were calculated from the spectrum of 2000 OP49 are shown in the upper right corner.  As in Figure \ref{fig.B1B2_Vtypes} multiple observations are connected and Vesta is plotted as the large filled circle. The HED regions are the same as in Figure \ref{fig.BAR_HED}. Magnya plots off the right edge of this figure with a BII center = 1.93 and BAR = 3.39. The BARs of the V-type asteroids tend to be much larger than those represented by the HEDs. 
} 
\label{fig.BAR_Vtypes}
\end{figure}

The band analysis of these inner belt V-types reveals a large offset in BAR between the asteroids and the HEDs (Fig. \ref{fig.BAR_Vtypes}). It is surprising that the difference in BARs is so pronounced: the mean asteroidal BAR is 2.28 while the mean meteoritic BAR is 1.58, a discrepancy much greater than the statistical errors associated with the telescopic data. The origin of this offset is unclear, however effects related to terrestrial weathering of the meteorites, differences between the asteroidal and laboratory environments (e.g.~temperature, grain size, phase angle), instrumental systematics, the technique of band analysis, compositional differences, impacts and space weathering are potential causes. In the following we argue that composition and space weathering are the only plausible explanations for this offset, though some combination could play a role.

{\it Weathering of Meteorites.} If terrestrial weathering of the meteorites caused them to have smaller BARs, then removing found meteorites  and only considering falls should reduce this offset. Of the 75 meteorite samples that were analyzed, 21 of these were observed as falls (Table \ref{tab.ap1.HEDs}). We find that removing found HEDs causes the regions in Figure \ref{fig.BAR_Vtypes} to shrink so that almost no overlap between the asteroids and the meteorites perists. Therefore it seems unlikely that terrestrial weathering is responsible for the discrepancy in BAR.

\begin{figure}[]
\begin{center}
\includegraphics[width=14cm]{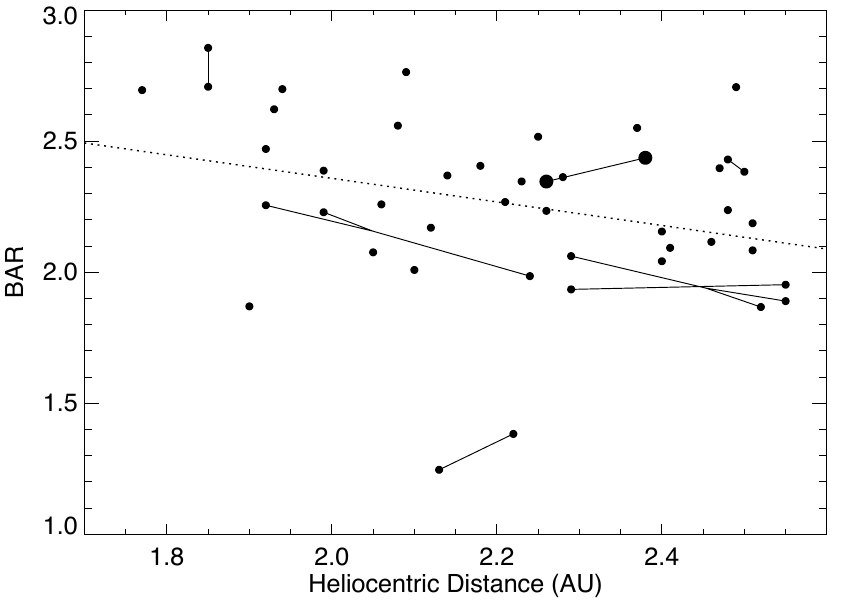}
\end{center}
\caption[Band area ratios of V-type asteroids versus heliocentric distance]{
Band area ratios versus heliocentric distance at the time of observation for our full set of V-type asteroids. Multiple observations are linked together. Vesta is plotted as the two large dots. This range of heliocentric distance corresponds to a temperature range of $\sim160-200$ K. This figure shows a weak inverse correlation between BAR and surface temperature. The best fit (based on chi-square minimization) is indicated by the dotted line and is described by Eqn. \ref{eqn.heliofit}.
} 
\label{fig.BAR_Helio}
\end{figure}

{\it Temperature Effects.} If the asteroid's low surface temperatures were the cause of their relatively large BARs, then heliocentric distance and BAR should be directly proportional. We plot these two quantities in Figure \ref{fig.BAR_Helio}. Although this figure shows a significant amount of scatter (which is likely due to observational issues and compositional effects), there is a weak inverse correlation between BAR and heliocentric distance ($r$). The best fit to this correlation is given by:
\begin{equation}
\mathrm{BAR} = 3.24 - 0.44 \left(\frac{r}{\mathrm{1~AU}}\right),
\label{eqn.heliofit}
\end{equation}
and can be interpreted as the mean spectral response of V-type mineralogies to changes in temperature. A decrease in BAR with temperature is supported by laboratory measurements of eucritic and diogenetic samples \citep{2002Icar..155..169H}. These experiments show that the width of Band II decreases with temperature faster than the width of Band I, thus causing BAR to drop. Therefore, the asteroidal data and these laboratory measurements suggest the opposite relationship than what would be expected if temperature were the cause of the offset between the asteroidal and meteoritic BARs. The combination of Equations \ref{eqn.temp} and \ref{eqn.heliofit} predicts a temperature-dependent BAR trend that can be tested by future measurements of HED meteorites or remote observation of V-type asteroids (particularly those made of Vesta by the Dawn spacecraft). 

It is unlikely that the increased thermal emission of the meteorites relative to the colder asteroids is a cause for the observed offset in BAR. Thermal emission from the meteorite samples causes an increase in their BII areas relative to those of the asteroids. This has the effect of producing larger BARs. If this thermal emission were accounted for then the discrepancy between the meteoritic and asteroidal BARs would increase.

{\it Grain Size Effects.} \citet{2005M&PS...40..445D} showed that the BARs of eucrites and diogenites increased with grain size. The particle sizes that they investigated ranged from 25 - 500 $\mu m$. They found that large grains produced BARs up to 1.9 for eucrites and a BAR of 2.04 was measured for a single diogenite. Neither of these values is high enough to reach the mean BAR of the V-types. Furthermore, the trend between BAR and grain size for the HEDs increases quickly up 100 $\mu m$ and then reaches saturation at larger grain sizes. If we ignore this saturation issue, then the large BARs of V-type asteroids might be explained by regoliths with particles beyond the upper end of the size range considered by \citet{2005M&PS...40..445D}. However, this contradicts the expected small regolith grain sizes ($\sim25~\mu m$) for V-type asteroids \citep{1994Metic..29..394H,1995Icar..115..374H} and the general result of polarization studies which suggest that asteroid surfaces have grains less than 300 $\mu m$ in size \citep{2002aste.conf..585C}. Furthermore, the multi-km asteroids in our sample should have regoliths with grains smaller than that observed on the sub-km asteroid Itokawa \citep{2009AJ....138.1557M}, which has particles down to at least 1 mm in size \citep{2006Sci...312.1330F}. We note that the BAR offset persists if we consider only RELAB samples with particles greater than or equal to 1 mm in size. For these reasons we argue that grain size effects are not sufficient to explain the observed BAR offset, however the presence of coarse regoliths on the surfaces of these asteroids could be a contributing factor.

{\it Phase Effects.} The asteroidal observations were performed at phase angles (defined as the angular separation between the Sun and observer as viewed from the target) between 2$^\circ$ and 30$^\circ$ (Table \ref{tab.obs}). As was the case for heliocentric distance, the BARs of the asteroids show significant scatter as a function of phase angle. However, our observations show a weak trend of increasing BAR with phase angle ($\phi$):
\begin{equation}
\mathrm{BAR} = 2.17 + 0.007 \left(\frac{\phi}{\mathrm{1~deg}}\right).
\label{eqn.phasefit}
\end{equation}
Telescopic observations of near-Earth V-type asteroids across a wide range of phase angles would be an important test for the validity of Equation \ref{eqn.phasefit}. Furthermore, future studies similar to those performed by \citet{2007LPI....38.1300C} will be essential to accurately characterize the effects of observing geometry on reflectance spectra of meteorite samples.

Very small ($< 2^\circ$) phase angles were not accessed by our telescopic observations, thus is is not clear whether the trend of Equation \ref{eqn.phasefit} remains linear or turns down for $\phi<2^\circ$. If it turns down then asteroidal phase angles of less than $2^\circ$ could result in small BARs similar to those of the HEDs. However, the effective phase angles (= angle of incidence + angle of emission) for the RELAB measurements were typically between 15$^\circ$ and 30$^\circ$ \citep{2001M&PS...36..761B,2005M&PS...40..445D}. Thus, if all else was equal, the BARs of the asteroids and the meteorites would have been comparable due to the overlapping range of observed phase angles. For this reason it seems unlikely that phase effects could have resulted in the observed BAR offset.

{\it Instrumental Systematics.} The offset in BAR could be due to an uncorrected systematic effect inherent to the SpeX instrument or the reduction process. However, this seems unlikely because the observations of 4 Vesta from 1981 (which were not performed with SpeX) result in a BAR (2.55) that is well outside of the range displayed by the meteorites and one that is similar to the value measured from the data of Vesta obtained with SpeX. Furthermore, reduction of SpeX data with different software packages \citep[e.g.][]{2003Icar..165..215K} and NIR observations of V-type asteroids with other instruments \citep[e.g.][]{2004Icar..171..120D} also result in large BARs. Future high-S/N observations with different NIR spectrographs will help to determine whether any systematics are affecting our results.

{\it Band Analysis Technique.} The definition of the red edge of Band II at 2.44 $\mu m$ acts to artificially reduce that band's area. It is apparent that the 2 $\mu m$ band for many of the asteroids extends past this cutoff (Fig. \ref{fig.obs}) and for some meteorites this band can extend out past 2.6 $\mu m$. Unfortunately the asteroidal data are sharply truncated at these longer wavelengths due to absorption by atmospheric water bands. Two tests have been performed which suggest that the BAR offset is not related to our somewhat arbitrary definition for the edge of the 2 $\mu m$ band.

First, we remeasured the BARs for all of the HEDs with the edge of BII defined at 2.6 $\mu m$. This redefinition had the effect of shifting the median BAR of the HEDs from 1.62 to 2.14. Although this reduced the discrepancy, the median BAR of the asteroids (equal to 2.26) with the cutoff at 2.44 $\mu m$ was still larger than this redefined meteoritic BAR. It is safe to assume that increasing the cutoff wavelength for the asteroids would also increase their BARs and thus preserve the offset relative to the HEDs.

\begin{figure}[]
\begin{center}
\includegraphics[width=14cm]{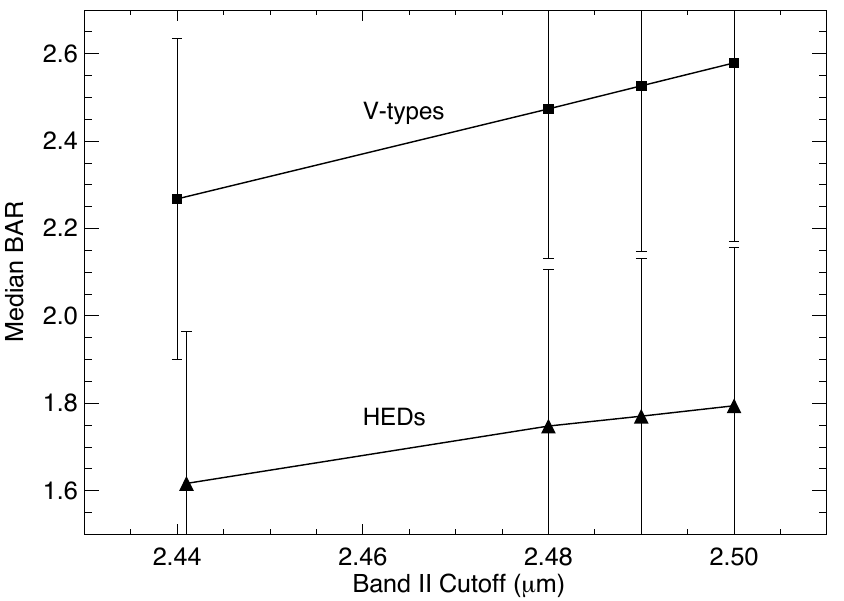}
\end{center}
\caption[Median BAR as a function of BII Cutoff]{
Median BAR for V-type asteroids and HED meteorites as a function of the cutoff wavelength for the 2 $\mu m$ absorption band. The vertical bars represent the standard deviations of the BARs. The data point at 2.44 $\mu m$ for the HEDs has been horizontally offset for clarity.
} 
\label{fig.cutoff}
\end{figure}

The effect of moving the edge of BII was tested for those asteroids with measured signal at wavelengths beyond 2.44 $\mu m$. Figure \ref{fig.cutoff} shows how the median BARs of the meteorites and the asteroids varies as a function of the BII cutoff.  All of the asteroids were included when the cutoff was defined at 2.44 $\mu m$, 39 asteroids have data long-wards of 2.48 $\mu m$, 31 asteroids could be measured with a cutoff at 2.49 $\mu m$, and 17 asteroids were included with a cutoff of 2.5 $\mu m$. Figure \ref{fig.cutoff} clearly shows that the increase in BAR associated with extending the edge of BII to longer wavelengths is similar for both the asteroids and the meteorites. This suggests that if the asteroid data could be extended to longer wavelengths, then the BAR offset from the meteorites would persist.

{\it Compositional Effects.} Compositional variation can not be ruled out as a cause for the difference between the BARs of the HEDs and the V-type asteroids. However, it would be surprising if the composition of the HEDs were wildly different from these V-type asteroids, particularly due to the numerous lines of evidence that suggest a direct link between these two populations \citep{1977GeCoA..41.1271C,1979aste.book..765D,1993Sci...260..186B,2008Icar..193...85N}. Making detailed statements regarding the mineralogical cause of such a large BAR offset between the HEDs and these V-types would require sophisticated spectral modeling \citep[e.g.][]{2007JGRE..11207005L,1999Icar..137..235S} and is beyond the scope of this work.

{\it Impact Effects.} The Vesta family is probably several billions of years old \citep{2005Icar..175..111B}, thus its members have experienced significant collisional evolution over their lifetimes. Impacts onto the surfaces of V-type asteroids would result in shock heating and the production of impact glass. The build-up of significant quantities of impact-generated material would affect reflectance spectra. We consider the results of two previous studies to investigate these effects, however further work should be done to better understand the spectroscopic implications of impacts on the surface of V-type asteroids.

\citet{2001M&PS...36..761B} included a spectrum of impact glass from the eucrite Macibini in their analysis of HED meteorites. A band analysis of this spectrum reveals a BAR of 1.77. This BAR is smaller than all but one of the V-type asteroids (2579 Spartacus), thus suggesting that the addition of impact glass to unshocked HED material would not result in the large BARs seen for asteroids.

\citet{1979LPI....10....1A} measured the effects of high pressure (597 kbar) shock on the reflectance spectrum of enstatite pyroxene. We found the BAR of their shocked and unshocked enstatite spectra equal to 2.09 and 2.36 respectively. Thus, the BAR of this sample decreased following shock. This result also suggests that impact-induced alteration can not produce the large BARs of V-type asteroids.

{\it Space Weathering Effects.} Space weathering (i.e.~modification of surfaces by micro-meteorite impact heating and irradiation by high energy particles) is a final possibility for explaining the large asteroidal BARs. Unfortunately little work has been done to understand the weathering of basaltic asteroids. This is partly due to the commonly held view that Vesta and the Vestoids have fresh or pristine surfaces  \citep{1995Icar..115..374H,2006IAUS..229..273P,2006A&A...451L..43V}; an assumption rooted in the spectroscopic similarity between V-type asteroids and the HEDs at visible wavelengths \citep{1970Sci...168.1445M,1977GeCoA..41.1271C,1993Sci...260..186B} and the assumption of lunar-style weathering \citep{2002aste.conf..585C} for asteroidal surfaces, irrespective of composition.

In general, the majority of studies on the space weathering of asteroids have focused on establishing a link between the visible wavelength spectral properties of S-type asteroids and ordinary chondrite meteorites \citep[e.g.][]{1988mess.book...35W,2002aste.conf..585C,2005Icar..173..132N,2008Icar..195..663W}. Space weathering on these bodies is assumed to be analogous to that which occurs on the Moon, i.e. the vaporization and redeposition of submicroscopic metallic iron by proton bombardment and/or micro-meteorite impact results in a depression of spectral absorption features and a reddening of spectral slope \citep{2002aste.conf..585C}.

Two independent sets of experiments have been performed to investigate the space weathering of HED meteorites. \citet{2006A&A...451L..43V} performed ion irradiation experiments to simulate weathering by the bombardment of solar wind particles. These authors irradiated a particulate sample (10-100 $\mu m$ grain size) of the eucrite Bereba at several different levels of ion fluence. We have performed a band analysis on the three samples (unaltered, moderately irradiated and heavily irradiated) shown in their Figure 1. The results of this analysis are summarized in Table \ref{tab.bereba}. Though the spectra of the irradiated samples, which exhibit steep red slopes and significantly depressed absorption features, do not resemble the spectra of any V-type asteroids, the band analysis does show that the BARs of the irradiated samples are larger by about 25\% relative to the unaltered Bereba sample. Qualitatively this shift is of the right magnitude and direction to potentially explain the BAR offset between the HEDs and V-type asteroids. It is worth noting that the band centers of the unaltered \citet{2006A&A...451L..43V} Bereba sample do not match those of the Bereba sample or any other of the eucrites from RELAB (Table \ref{tab.ap1.HEDs}). This is likely due to differences in the preparation of the samples: RELAB uses a horizontally mounted sample tray, whereas \citet{2006A&A...451L..43V} used a compacted, vertically mounted sample. Nevertheless, their results are compelling and certainly merit further investigation.

\begin{figure}[]
\begin{center}
\includegraphics[width=14cm]{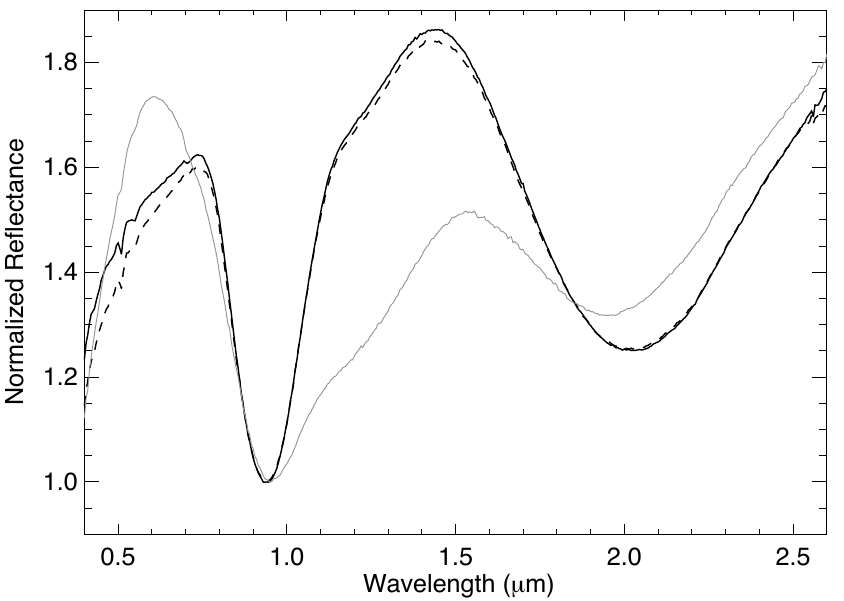}
\end{center}
\caption[Spectra of eucrite Millbillillie from the \citet{1998LPI....29.1940W} laser irradiation experiment]{
Reflectance spectra of eucrite Millbillillie from the study of \citet{1998LPI....29.1940W}. The spectra have been normalized by the reflectance at the wavelengths of their respective Band I centers (Table \ref{tab.laser}) to emphasize the change in band parameters with irradiation level. The black line is the unaltered particulate sample (grain size $<75~\mu m$). The dashed line is the partially irradiated material (grain size $<75~\mu m$). The grey line is the spectrum of the fully irradiated material (grain size $>75~\mu m$).
} 
\label{fig.laser}
\end{figure}

The second set of HED weathering experiments were performed by \citet{1997LPI....28.1505W,1998LPI....29.1940W} who used laser irradiation to simulate weathering by micro-meteorite impacts. Although one eucrite (Millbillillie) and one diogenite (Johnstown) were originally included in this study, problems related to the intensity of the laser necessitated a second series of experiments in which only Millbillillie was studied. We will only discuss the results of this second set of experiments. Prior to irradiation Millbillillie was ground to particle sizes of  $<75~\mu m$. It was found that irradiation produced spherical blobs of glassy material that were similar in composition to the original sample and were typically $>75~\mu m$ in diameter. These investigators divided their irradiation products into two categories (fully and partially irradiated) delineated by a cutoff in particle size at $75~\mu m$. The fully irradiated products ($>75~\mu m$ in size) contained $\sim95\%$ of the glassy material produced by the laser irradiation, the partially irradiated material contained the remaining 5\%. We have performed our band analysis on the unaltered, partially and fully irradiated spectra (Figure \ref{fig.laser}) from the \citet{1998LPI....29.1940W} study. The results of this band analysis are given in Table \ref{tab.laser}. 

The spectrum of the fully irradiated material (Figure \ref{fig.laser}) does not resemble any of the V-type asteroids that were included in our survey (Figure \ref{fig.obs}). Furthermore, it does not have band centers or a BAR that are comparable to any of the meteorites and asteroids that were analyzed. This suggests that the fully irradiated sample is not a good analog for space weathered V-type asteroids, which is not terribly surprising due to the large particle sizes of this material. On the other hand the partially irradiated sample closely resembles unaltered Millbillillie material and has band centers that are typical for both the HEDs and V-type asteroids. More interesting though is that the partially irradiated sample has a BAR that is greater by $\sim2\%$ (due to a greater decrease in the Band I area relative to Band II, see Fig. \ref{fig.laser}). This small fraction is far from explaining the $\sim40\%$ offset between the BARs of the  HEDs and V-type asteroids, however it again emphasizes the need for further study of space weathering processes on V-type asteroids. 

\begin{figure}[]
\begin{center}
\includegraphics[width=14cm]{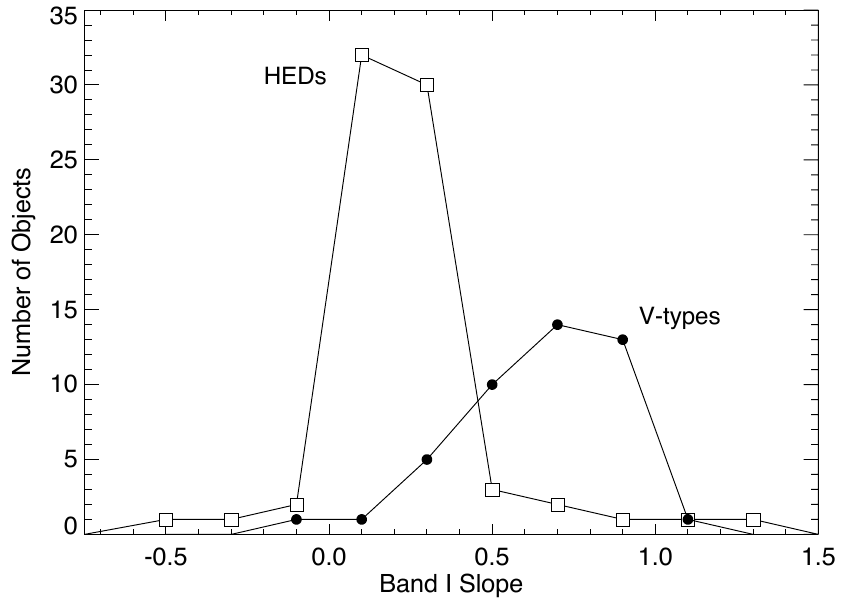}
\end{center}
\caption[Histograms of Band I Slope]{
Histograms of Band I slope for V-type asteroids and HED meteorites. The asteroids clearly have larger Band I slopes, consistent with the expected spectroscopic effects from the production of submicroscopic metallic iron by space weathering.
} 
\label{fig.slope}
\end{figure}

\begin{figure}[]
\begin{center}
\includegraphics[width=14cm]{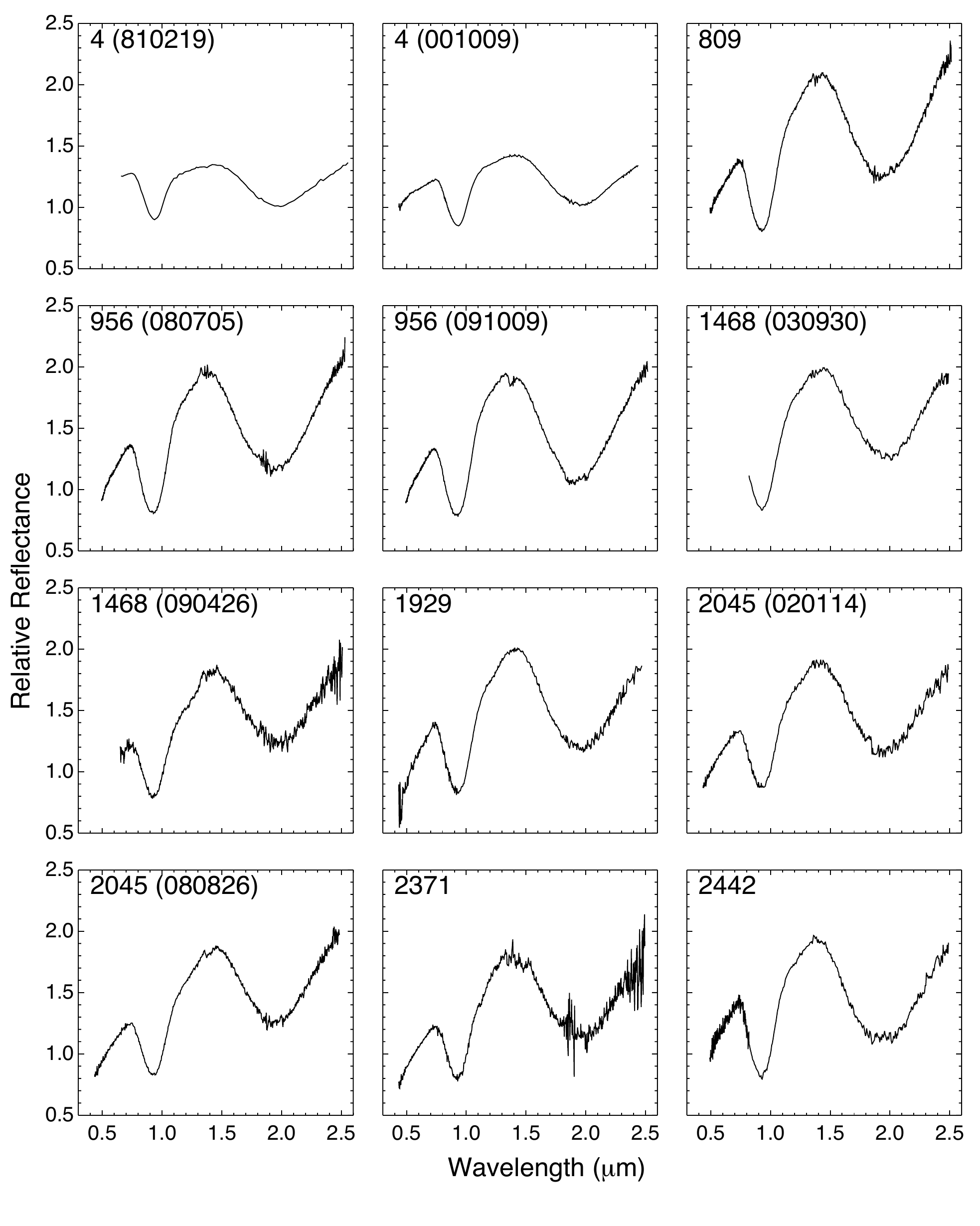}
\end{center}
\caption[NIR Spectra of V-type Asteroids]{
} 
\label{fig.obs}
\label{lastfig}
\end{figure}

If V-type asteroids have weathered surfaces containing submicroscopic metallic iron, then their continuum slopes should be steeper that those of the HEDs \citep{2002aste.conf..585C}. We check this by comparing Band I slopes (Figure \ref{fig.slope}). This figure shows that the V-types tend to have larger Band I slopes than the HEDs\footnote{Note that the ratio of the reflectance peaks at $\sim$1.4 $\mu m$ and $\sim$0.75 $\mu m$ as defined by \citet{1986JGR....9111641C} is similarly indicative of continuum slope and shows the same trend as that of the Band I slopes in Figure \ref{fig.slope}.}. Though this is consistent with the expected spectroscopic effects induced by space weathering, it is unclear why V-type asteroids still have pronounced 1 and 2 $\mu m$ absorption features while weathered lunar basalts and S-type asteroids exhibit reduced band depths \citep{2002aste.conf..585C}. This issue may be related to compositional differences: the spectroscopic manifestation of weathering on V-type asteroids may be different because of their lower iron and olivine abundances \citep{2001M&PS...36..761B} relative to both the ordinary chondrites and typical lunar material \citep{1998psc..book.....L}.

Expanded irradiation and proton bombardment experiments with howardite samples \citep[which are the best analogs to Vesta's spectrum,][]{2006IAUS..229..273P} could offer insight on the spectral implications of weathering on Vesta-like surfaces. If the observed spectroscopic differences between the V-type asteroids and the HED meteorites are due to space weathering, then older, more weathered regions on Vesta's surface should have band centers similar to the HEDs, but BARs that are significantly larger. The Visible and Infrared Mapping Spectrometer (VIR-MS) onboard the Dawn spacecraft \citep{2004P&SS...52..465R} will help to provide insight on this issue.

{\it A Combination of Effects.} The surfaces of asteroids are complex environments that differ from laboratory samples of meteorites in multiple ways. In this section, we have shown that terrestrial weathering of the HEDs, temperature effects, phase effects, and impacts all act to increase the BAR discrepancy between the asteroids and the meteorites. Thus it is reasonable to suggest that a combination of these effects would also increase the BAR discrepancy. Furthermore, we have shown that the technique of band analysis and instrumental systematics do not have a significant impact on producing discrepant BARs. This leaves grain size, composition and space weathering to explain the asteroidal BARs. Although some of these cannot individually explain the BAR offset, further work is required to understand the spectroscopic effects from a combination of these factors.

\section{Band Parameters and Orbital Properties}

\subsection{Comparison to \citet{1998AMR....11..163H}}

\citet{1998AMR....11..163H} found that V-type asteroids outside of the Vesta family tended to have steeper spectral slopes at visible wavelengths than V-types within the family, a trend that was attributed to an enhancement of weathered material on the surfaces of non-family members. At the time of their study the dynamical boundaries of the Vesta family extended to $\Delta v \sim 0.5$ km/s. However, more recent definitions of the family \citep[][]{2008Icar..193...85N} now incorporate all of the \citet{1998AMR....11..163H} asteroids and include members out to $\Delta v \sim 1$ km/s. Thus, it is surprising that this spectro-dynamical trend exists for V-types that can all be linked to the same collisional origin.

\citet{1998AMR....11..163H} measured ``visible redness"  (roughly analogous to $V-R$ color)  and Band I depth (defined as the difference in reflectance between the peak at $\sim0.75~\mu m$ and the Band I minimum) for each of their targets using the Modified Gaussian Model \citep[MGM,][]{1990JGR....95.6955S}. Ideally, we would do the same for all 39 asteroids observed here, however application of the MGM to such a large data set is impractical. Furthermore, our spectra do not include visible wavelength data for each asteroid, thus precluding the measurement of visible redness for some objects. However, it was shown in the previous section that the difference in Band I slope between V-type asteroids and HED meteorites may be linked to the effects of space weathering (Fig. \ref{fig.slope}). In addition, it is believed that the production of submicroscopic metallic iron via space weathering has spectral implications that are in some ways similar at visible and NIR wavelengths \citep{2002aste.conf..585C}. Therefore we use Band I slope as a proxy for visible redness.

The results of  \citet{1998AMR....11..163H} suggest that V-types with $\Delta v > 0.5$ km/s should tend to have large Band I slopes. The mean Band I slope for V-types with $\Delta v > 0.5$ km/s is 0.79 while the mean for V-types with $\Delta v < 0.5$ km/s is 0.70, a shift of approximately 10\%. This trend is consistent with the \citet{1998AMR....11..163H} result, however it is not as pronounced: the objects that they referred to as non-Vesta family members showed $\sim20\%$ higher mean visible redness than objects within the Vesta family. 

Though this shift in Band I slope is consistent with \citet{1998AMR....11..163H}, there are several reasons to suggest that it is not a significant result. First, the probability associated with the two-sided KS statistic suggests that the distributions of Band I slopes for V-types with $\Delta v$ on either side of 0.5 km/s are unlikely ($<1\sigma$) to be derived from distinct parent populations. Second, variations in Band I slope of $\pm0.1$ (similar in magnitude to the 10\% offset) are seen with multiple observations of the same asteroid (Table \ref{tab.band_params}). Finally, the subset of 28  V-types with $\Delta v>0.5$ km/s have Band I slopes distributed across the full range of values shown by the HEDs and other V-types, with an excess of only a few objects at large values. This is in contrast to  \citet{1998AMR....11..163H} who show that 7 out of 8 of their non-Vesta family members have visible redness greater than that of the HEDs and other V-types. For these reasons we interpret the difference in Band I slopes for V-types with $\Delta v$ on either side of 0.5 km/s as a consequence of low number statistics.

Our inability to convincingly reproduce at NIR wavlengths the results of \citet{1998AMR....11..163H} may be related to the assumed equivalence between visible redness and Band I slope. It could also be related to the sensitivity of NIR spectra to slope variations, particularly at the endpoints of the spectra. Unfortunately, only seven of our targets (4 Vesta, 1929 Kollaa, 2442 Corbett, 3155 Lee, 3657 Ermolova, 4038 Kristina, and 4215 Kamo) are shared in common with \citet{1998AMR....11..163H}, and of these seven objects none have $\Delta v>0.5$ km/s. NIR observations of the V-type asteroids that \citet{1998AMR....11..163H} categorize as ``non Vesta family" objects are needed to better address this issue. 

\subsection{Spectral Implications of \citet{2008Icar..193...85N}}

The dynamical simulations performed by \citet{2008Icar..193...85N} did not produce a sufficient number of Vestoid fragments with low inclination orbits to explain the observed distribution of V-type asteroids in the inner Main Belt. They concluded that the inner Main Belt may contain V-type asteroids from non-Vestoid parent bodies or fragments from the surface of Vesta that were ejected at an epoch earlier than the main family-forming collision. In either case it is reasonable to suggest that the regions of orbital element space under-populated in the  \citet{2008Icar..193...85N} simulations may contain V-types that are spectroscopically distinct from ``ordinary" Vesta family members.

The under-populated region in the \citet{2008Icar..193...85N} simulations was referred to as Cell 2 and defined as: $2.32 < a < 2.48$, $0.05<e<0.2$, $2^\circ<i<6^\circ$. This region was chosen to be diagnostic of the number of V-types at low inclinations. Table \ref{tab.nesvorny} lists the median band centers and BARs for V-type asteroids in four different dynamical regions: inside Cell 2, outside of Cell 2, inclinations less than $6^\circ$ and inclinations greater than $6^\circ$. Note that all but three of the objects studied here have orbital eccentricities between 0.05 and 0.2, thus the region labelled as ``Outside Cell 2" in Table \ref{tab.nesvorny} actually includes those eccentricities spanned by Cell 2. This is done to increase the number statistics within that region and does not affect any of the following conclusions. Eleven objects are included in Cell 2, 33 objects orbit outside of Cell 2, 22 have inclinations less than $6^\circ$ and 23 have inclinations greater than $6^\circ$.

The median values of the band centers for objects in Cell 2 and with $i<6^\circ$ are at longer wavelengths than the band centers for objects outside of these regions (Table \ref{tab.nesvorny}). This shift to longer wavelengths is larger than the maximum statistical error bars on the band centers and is most pronounced in the comparison of objects on either side of $i=6^\circ$ (the apparent difference in BAR is smaller than the statistical errors and thus may not be significant). It is important to note that this shift is subtle and may simply be a consequence of low number statistics. Nevertheless, it is interesting that the dynamical simulations of  \citet{2008Icar..193...85N} may have an observational counterpart.

We can speculate about what it means for the interpretation of V-type asteroids in the inner Main Belt if this band center-inclination trend is real. Amongst the HEDs, the relatively Ca- and Fe-rich eucrites have band centers at longer wavelengths than the Mg-rich and Fe-poor diogenites \citep{2001M&PS...36..761B}. Thus, V-type asteroids at low inclinations may be characterized as generally eucritic in composition. If these objects represent an older population of Vestoids \citep{2008Icar..193...85N}, then they may have come from a crater other than the one on Vesta's south pole \citep[which is thought to have excavated down into Vesta's diognite-like upper mantle and produced the main Vesta family;][]{1997Icar..127..130G}. Two depressions in Vesta's northern hemisphere, each approximately 150 km in diameter, have been interpreted as impact craters \citep{1997Sci...277.1492T}. These craters may not have excavated down to Vesta's upper mantle and thus would generate fragments primarily eucritic in composition. Low-inclination V-type asteroids, with eucrite-like band centers, could have originated from one of these smaller craters. If these speculations are correct, then we expect that the Dawn spacecraft \citep{2004P&SS...52..465R} will not observe large outcroppings or concentrations of diogenite-like material at the bottom of the two $\sim150$ km-diameter craters.

\section{Summary}

We have observed and analyzed the spectra of 39 V-type asteroids. Comparison of their band parameters to those of HED meteorites from the RELAB database reveals a close correlation between band centers. We do not find the wide range of band centers that was reported by \citet{2004Icar..171..120D}. We suspect that this difference is due to the lower S/N of the \citet{2004Icar..171..120D} data set. We were able to confirm an offset in BAR between the HEDs and V-type asteroids \citep{2005M&PS...40..445D} and argue that this offset is consistent with our initial understanding of space weathering effects on Vesta-like mineralogies However, further work is necessary to understand whether  a combination of grain size, composition and/or weathering could be responsible for this offset. Several other possible causes were discussed and found to be unlikely.  We were unable to reproduce at NIR wavelengths the spectro-dynamical association found by \citet{1998AMR....11..163H}, namely inner belt V-types do not show any correlation between slope across their 1 $\mu m$ bands and distance from Vesta in orbital element space. We note that a search for correlation between other spectroscopic characteristics (band centers, slopes, depths and areas) and dynamical properties (semi-major axis, inclination, eccentricity, $\Delta v$) did not reveal any significant results. 

We did not find any new evidence to suggest the presence of V-type asteroids with non-Vesta mineralogies [e.g.~1459 Magnya and 21238 (1995 WV7)]. Instead, the band parameters of these objects seem to represent a continuum of compositions that are consistent with an origin from a single parent body, most likely 4 Vesta. Only asteroid 2579 Spartacus is found to have a band area ratio that is significantly offset from the general trend represented by the other targets. This has been noted before \citep{2001M&PS...36..761B} and could be due to its origin on another parent body, however it could also be a large fragment that originated from greater depths (relative to the other targets) within the Vesta parent body. The lack of additional spectroscopic outliers amongst the V-type asteroids in the inner Main Belt implies that they are of a common origin. However, this does not preclude the possibility that inner belt V-types include basaltic crustal fragments from multiple differentiated parent bodies that are indistinguishable with band analysis techniques. 

We have reported that V-type asteroids with low inclinations ($i<6^\circ$) in the inner Main Belt tend to have band centers shifted to longer wavelengths. This is compelling in light of the dynamical results of \citet{2008Icar..193...85N}, however additional data should be obtained to confirm or refute the significance of this finding. In particular, a focused study on low-inclination V-type asteroids may provide further insight on whether these objects are spectroscopically distinct.

This study has resulted in several predictions that can be tested by the Dawn spacecraft when it enters into orbit around Vesta in the summer of 2011. First, we have parameterized the dependence between temperature and BAR (Equations \ref{eqn.temp} and \ref{eqn.heliofit}). Spectroscopic observations with VIR-MS of regions at various temperatures on Vesta's surface can be used check this dependence. We have also suggested that fresher, less weathered surfaces on Vesta (e.g. impact craters) should have smaller BARs than the surrounding terrain. And finally, if the low inclination V-type asteroids in our study are predominantly eucritic in composition and were removed from one of the minor impact craters in Vesta's northern hemisphere \citep{1997Sci...277.1492T}, then we expect that these craters will be devoid of significant quantities of diogenite-like material.

\section*{Appendix A: Asteroid Spectra }

In this appendix we present the spectra of all V-type asteroids analyzed in this study. Numerical designations are shown in each panel. Some targets were observed on multiple occasions: the dates of these observations are indicated in a year, month, day format (YYMMDD) following the object designation. Observations prior to the year 2006 were either published in \citet{2009Icar..202..160D}, obtained as part of the ongoing MIT-UH-IRTF Joint Campaign for NEO Spectral Reconnaissance (NEOSR), or for the spectrum of 4 Vesta from 1981, obtained by \citet{1997Icar..127..130G}. When available, visible wavelength data from one of several sources \citep{2002Icar..158..106B,2004Icar..172..179L,2008Icar..198...77M} have been appended to the NIR data. Multiple NEOSR observations of 2653 Principia, 2851 Harbin, 3155 Lee, 3782 Celle and 4215 Kamo were combined into composite spectra. All spectra have been normalized to unity at 1 $\mu m$ and are all presented at the same scale.

(FIGURE \ref{fig.obs} SHOULD GO HERE)

\section*{Appendix B: HED Band Parameters}

This appendix contains a table of the measured HED band parameters. The columns in this table are: name of the sample, meteorite type, mean particle size of the sample when the spectra were measured, BI and BII centers, and the ratio of BII area to BI area (BAR). The samples are listed alphabetically within each of the HED subgroups. Some samples were not ground into particles to measure their reflectance spectrum: these are indicated as ``chip" in the particle size column.

(Table \ref{tab.ap1.HEDs} SHOULD GO HERE)

\ack
This work has benefitted from the generosity of several individuals who have kindly shared the results of previous investigations. Thanks to Michael Gaffey for providing his observations of Vesta, to Paul Hardersen for his data of Magnya, and to Pierre Vernazza for providing the results of his HED weathering experiments. We are grateful for the careful reviews and insightful comments from Ed Cloutis and an anonymous reviewer. Thanks to Robert Jedicke, Eric Gaidos and Scott Sheppard for helpful comments on various drafts of this manuscript. This work has used spectra obtained by several investigators at the NASA RELAB facility at Brown University. We would like to acknowledge the RELAB database as a vital resource to the planetary astronomy community. N.M. would like to acknowledge the support of NASA GSRP grant NNX06AI30H. Part of the data utilized in this publication were obtained and made available by the MIT-UH-IRTF Joint Campaign for NEO Reconnaissance. The IRTF is operated by the University of Hawaii under Cooperative Agreement no. NNX-08AE38A with the National Aeronautics and Space Administration, Science Mission Directorate, Planetary Astronomy Program.. The MIT component of this work is supported by the National Science Foundation under Grant No. 0506716. We wish to recognize and acknowledge the very significant cultural role and reverence that the summit of Mauna Kea has always had within the indigenous Hawaiian community.  We are most fortunate to have the opportunity to conduct observations from this mountain.

\label{lastpage}





\clearpage	

\begin{center}
\begin{longtable}{llcccc}
\hline 
\hline
& & Heliocentric & & \\
Object & UT Date & Distance (AU) & Phase (deg) & Mag. & $t_{exp}$ (min) \\
\hline
4 Vesta				& Feb. 18-20, 1981$^a$	& 2.38	& 4.0		& 6.12 & -- \\
					& Oct. 9, 2000			& 2.26	& 25.9	& 7.3 & 10 \\
809 Lundia			& Aug. 26, 2008		& 1.93	&23.9	& 14.6 & 8 \\
956 Elisa				& Jul. 5, 2008			& 1.85	& 16.6	& 14.6 & 16 \\
					& Oct. 9, 2008			& 1.85	& 16.9	& 15.7 & 88 \\
1468 Zomba			& Sep. 30, 2003		& 1.60	& 38.3	& & \\
					& Apr. 26, 2009			& 2.51	& 13.0	& 17.4 & 80 \\
1929 Kollaa			& Feb. 19, 2001		& 2.21	& 14.4	& 15.3 & 24 \\
2045 Peking			& Jan. 14, 2002		& 2.48	& 21.7	& 16.7 & 40 \\
					& Aug. 26, 2008		& 2.50	& 16.9	& 16.2 & 87 \\
2371 Dimitrov			& Aug. 1, 2009			& 2.47	& 19.6	& 16.7 & 40 \\
2442 Corbett			& Sep. 30, 2003		& 2.25	& 7.8		& & \\
2511 Patterson			& May 7, 2004			& 2.23	& 21.2	& 16.1 & 32 \\
2566 Kirghizia			& May 8, 2002			& 2.41	& 9.8		& 16.0 & 32 \\
2579 Spartacus		& Oct. 10, 2000			& 2.22	& 18.9	& 16.4 & 28 \\
2653 Principia			& Nov. 26, 2002		& 2.53	& 17.8	& 16.3 & 48 \\
					& Jul. 16, 2005			& 2.59	& 3.1		& 15.5 & 28 \\
					& Apr. 19, 2008			& 2.29	& 12.3	& 15.3 & 12 \\
2763 Jeans			& Jun. 26, 2004		& 2.24	& 12.6	& 15.7 & 32 \\
					& Jul. 5, 2008			& 1.99	&12.2	& 14.9 & 16 \\
					& Aug. 26, 2008		& 1.92	&17.3	& 14.9 & 24 \\
2795 Lepage			& Apr. 9, 2005			& 2.26	& 4.7		& 15.9 & 24 \\
2823 van der Laan		& Nov. 22, 2005		& 2.20	& 16.1	& & \\
2851 Harbin			& Aug. 24, 2001		& 2.42	& 8.7		& 15.6 & 28 \\
					& Jan. 12, 2003		& 2.20	& 16.0	& 15.9 & 48 \\
2912 Lapalma			& Feb. 20, 2001		& 2.14	& 8.7		& 15.3 & 32 \\
3155 Lee				& Jun. 22, 2001		& 2.56	& 8.4		& 16.2 & 28 \\
					& Jul. 14, 2005			& 2.43	& 7.9		& 15.9 & 28 \\
3657 Ermolova			& Aug. 1, 2009			& 2.18	& 25.4	& 16.6 & 40 \\
3703 Volkonskaya		& Jun. 3, 2006			& 2.25	& 9.8		& & \\
3782 Celle			& Nov. 26, 2002		& 2.60	& 17.1	& 16.8 & 32 \\
					& Jun. 25, 2004		& 2.19	&13.3	& 15.5 & 28 \\
4038 Kristina			& Oct. 28, 2002			& 2.08	& 10.3	& & \\
4188 Kitezh			& Aug. 14, 2001		& 2.10	& 10.8	& 15.3 & 28 \\
4215 Kamo			& Nov. 11, 2002		& 2.52	& 17.3	& 16.5 & 64 \\
					& Jul. 15, 2005			& 2.51	& 3.2		& 15.6 & 40 \\
					& Apr. 19, 2008			& 2.29	& 9.5		& 15.4 & 12 \\
					& Aug. 1, 2009			& 2.55	& 7.8		& 15.9 & 20 \\
4796 Lewis			& Jan. 9, 2009			& 2.46	& 13.9	& 17.0 & 33 \\
5111 Jacliff			& Sep. 5, 2005			& 2.06	& 7.6		& & \\
5481 Kiuchi			& Aug. 1, 2009			& 2.37	& 10.5	& 16.3 & 54 \\
5498 Gustafsson		& Aug. 26, 2008		& 1.92	& 7.8		& 16.3 & 93 \\
7800 Zhongkeyuan		& Jan. 9, 2009			& 2.51	& 2.8		& 17.1 & 27 \\
9481 Menchu			& Aug. 26, 2008		& 2.48	& 6.1		& 17.3 & 40 \\
9553 Colas			& Jan. 8, 2009			& 1.99	& 20.3	& 17.4 & 120 \\
16416 (1987 SM3)		& Nov. 23, 2007		& 2.40	& 6.0		& 16.9 & 64 \\
26886 (1994 TJ2)		& Jul. 5, 2008			& 2.09	& 10.9	& 17.2 & 40 \\
27343 (2000 CT102)	& Aug. 26, 2008		& 1.94	& 11.7	& 16.6 & 48 \\
33881 (2000 JK66)		& Nov. 23, 2007		& 1.77	& 23.3	& 16.6 & 48 \\
36412 (2000 OP49)		& Nov. 23, 2007		& 2.12	& 11.8	& 16.9 & 40 \\
38070 (1999 GG2)		& Oct. 5, 2006			& 1.90	& 4.1		& 16.9 & 32 \\
50098 (2000 AG98)		& Sep. 4, 2005			& 1.94	& 11.4	& & \\
97276 (1999 XC143)	& Nov. 23, 2007		& 2.05	& 2.4		& 16.6 & 56 \\
\hline
\caption[]{Summary of NIR Spectroscopic Observations\\
The columns in this table are: object number and designation, UT date of observation, heliocentric distance at the time of observation, the phase angle of the target, V-band magnitude of the target from the JPL HORIZONS system on the date of observation and net exposure time in minutes.\\
$^a$Data from \citet{1997Icar..127..130G}
}
\label{tab.obs}
\end{longtable}
\end{center}

\clearpage

\begin{center}
\scriptsize
\begin{longtable}{lccccccc}
\hline 
\hline
  & Heliocentric & Mean Surface & BI Center & BI & $\Delta$BII & BII Center & \\
Object & Distance (AU) & Temp. (K) & ($\mu$m) & Slope &  ($\mu$m) & ($\mu$m)  & BAR \\
\hline
      4 Vesta & 2.26 & 173.9 & 0.935 & 0.311 & 0.022 & 1.986 & 2.346 \\
             & 2.38 & 169.4 & 0.943 & 0.090 & 0.023 & 2.003 & 2.436 \\
     809 Lundia & 1.93 & 188.1 & 0.937 & 1.093 & 0.020 & 1.954 & 2.622 \\
      956 Elisa & 1.85 & 192.2 & 0.932 & 0.963 & 0.019 & 1.964 & 2.707 \\
             & 1.85 & 192.2 & 0.931 & 1.013 & 0.019 & 1.940 & 2.856 \\
      1468 Zomba & 1.60 & 206.6 & ----- & ----- & 0.016 & 1.992 & ------ \\
             & 2.51 & 165.0 & 0.938 & 0.902 & 0.024 & 2.001 & 2.187 \\
     1929 Kollaa & 2.21 & 175.8 & 0.941 & 0.942 & 0.022 & 1.980 & 2.268 \\
     2045 Peking & 2.48 & 166.0 & 0.938 & 0.861 & 0.024 & 1.976 & 2.430 \\
             & 2.50 & 165.3 & 0.938 & 0.889 & 0.024 & 1.968 & 2.383 \\
   2371 Dimitrov & 2.47 & 166.3 & 0.936 & 0.968 & 0.024 & 2.008 & 2.397 \\
    2442 Corbett & 2.25 & 174.2 & 0.933 & 0.812 & 0.022 & 1.962 & 2.517 \\
  2511 Patterson & 2.23 & 175.0 & 0.933 & 0.785 & 0.022 & 1.977 & 2.346 \\
  2566 Kirghizia & 2.41 & 168.4 & 0.939 & 0.583 & 0.023 & 1.989 & 2.093 \\
  2579 Spartacus & 2.22 & 175.4 & 0.935 & 0.516 & 0.022 & 2.011 & 1.383 \\
             & 2.13 & 179.1 & 0.952 & 0.907 & 0.021 & 2.045 & 1.246 \\
  2653 Principia & 2.55$^a$ & 163.7 & 0.932 & 0.608 & 0.024 & 1.994 & 1.952 \\
             & 2.29 & 172.7 & 0.940 & 0.742 & 0.022 & 2.012 & 1.934 \\
      2763 Jeans & 2.24 & 174.6 & 0.944 & 0.664 & 0.022 & 2.019 & 1.985 \\
             & 1.99 & 185.3 & 0.941 & 0.792 & 0.020 & 1.994 & 2.229 \\
             & 1.92 & 188.6 & 0.942 & 0.898 & 0.020 & 2.027 & 2.255 \\
     2795 Lepage & 2.26 & 173.9 & 0.938 & 0.532 & 0.022 & 1.973 & 2.234 \\
2823 van der Laan & 2.20 & 176.2 & ----- & ----- & 0.022 & 1.989 & ------ \\
     2851 Harbin & 2.28$^a$ & 173.1 & 0.922 & 0.693 & 0.022 & 1.934 & 2.363 \\
    2912 Lapalma & 2.14 & 178.7 & 0.930 & 0.930 & 0.021 & 1.951 & 2.369 \\
        3155 Lee & 2.49$^a$ & 165.6 & 0.919 & 0.823 & 0.024 & 1.933 & 2.706 \\
   3657 Ermolova & 2.18 & 177.0 & 0.931 & 0.703 & 0.022 & 1.942 & 2.406 \\
3703 Volkonskaya & 2.25 & 174.2 & ----- & ----- & 0.022 & 1.964 & ------ \\
      3782 Celle & 2.40$^a$ & 168.7 & 0.935 & 0.546 & 0.023 & 1.966 & 2.042 \\
   4038 Kristina & 2.08 & 181.2 & 0.919 & 0.608 & 0.021 & 1.984 & 2.559 \\
     4188 Kitezh & 2.10 & 180.4 & 0.939 & 0.777 & 0.021 & 1.979 & 2.009 \\
       4215 Kamo & 2.52$^a$ & 164.6 & 0.925 & 0.337 & 0.024 & 1.988 & 1.867 \\
             & 2.29 & 172.7 & 0.932 & 0.372 & 0.022 & 1.975 & 2.061 \\
             & 2.55 & 163.7 & 0.926 & 0.629 & 0.024 & 1.977 & 1.890 \\
      4796 Lewis & 2.46 & 166.6 & 0.935 & 0.732 & 0.024 & 1.968 & 2.116 \\
    5111 Jacliff & 2.06 & 182.1 & 0.926 & 0.558 & 0.021 & 1.964 & 2.259 \\
     5481 Kiuchi & 2.37 & 169.8 & 0.932 & 0.922 & 0.023 & 1.959 & 2.550 \\
 5498 Gustafsson & 1.92 & 188.6 & 0.941 & 0.986 & 0.020 & 1.990 & 2.470 \\
7800 Zhongkeyuan & 2.51 & 165.0 & 0.939 & 0.902 & 0.024 & 1.962 & 2.083 \\
     9481 Menchu & 2.48 & 166.0 & 0.940 & 0.296 & 0.024 & 1.958 & 2.237 \\
      9553 Colas & 1.99 & 185.3 & 0.932 & 0.803 & 0.020 & 1.949 & 2.387 \\
    16416 (1987 SM3) & 2.40 & 168.7 & 0.939 & 0.887 & 0.023 & 1.988 & 2.155 \\
    26886 (1994 TJ2) & 2.09 & 180.8 & 0.922 & 0.877 & 0.021 & 1.923 & 2.763 \\
  27343 (2000 CT102) & 1.94 & 187.7 & 0.925 & 1.021 & 0.020 & 1.935 & 2.698 \\
   33881 (2000 JK66) & 1.77 & 196.5 & 0.935 & 1.105 & 0.018 & 1.948 & 2.695 \\
   36412 (2000 OP49) & 2.12 & 179.5 & 0.942 & 0.999 & 0.021 & 1.986 & 2.170 \\
    38070 (1999 GG2) & 1.90 & 189.6 & 0.941 & 0.377 & 0.019 & 1.988 & 1.870 \\
   50098 (1999 AG98) & 1.94 & 187.6 & ----- & ----- & 0.019 & 1.968 & ------ \\
  97276 (1999 XC143) & 2.05 & 182.5 & 0.946 & 0.370 & 0.021 & 2.037 & 2.076 \\
\hline
\caption[]{Band Parameters of V-type Asteroids\\
The columns in this table are: object designation, heliocentric distance at the time of observation, estimated mean surface temperature (Eqn. \ref{eqn.temp}), BI center, BI depth, temperature correction to BII center, BII center and the BII to BI area ratio (BAR). At the resolution of the spectra, the temperature correction for all of the BI centers is 0.002 $\mu$m. The band centers in this table have been temperature corrected. The maximum statistical errors (as determined from the spectrum of asteroid 2000 OP49) for these band parameters are: $\sigma_{BI}=0.003$, $\sigma_{BII}=0.011$ and $\sigma_{BAR}=0.23$.\\
$^a$Represents an average heliocentric distance from two separate observations.
}
\label{tab.band_params}
\end{longtable}
\end{center}

\clearpage

\begin{table}
\begin{center}
\begin{tabular}{lccccc}
\hline 
\hline
Sample Info & BI Center ($\mu$m) & BI Area & BII Center ($\mu$m) & BII Area & BAR \\
\hline
Unaltered particulate & 0.928 & 0.036 & 2.031 & 0.063 & 1.749\\
Moderately irradiated & 0.934 & 0.026 & 2.015 & 0.058 & 2.188\\
Heavily irradiated & 0.944 & 0.021 & 1.751 & 0.043 & 2.044\\
\hline
\end{tabular}
\caption[Band Parameters for Eucrite Bereba]{Band Parameters for Eucrite Bereba}
\label{tab.bereba}
\end{center}
\end{table}

\clearpage

\begin{table}
\begin{center}
\begin{tabular}{lccccc}
\hline 
\hline
Sample Info & BI Center ($\mu$m) & BI Area & BII Center ($\mu$m) & BII Area & BAR \\
\hline
Unaltered particulate & 0.942 & 0.151 & 2.025 & 0.234 & 1.553\\
\hspace{0.1cm} (grain size  $<75~\mu$m) & & & & &\\
Partially irradiated & 0.942 & 0.144 & 2.023 & 0.229 & 1.586\\
\hspace{0.1cm} (grain size  $<75~\mu$m) & & & & &\\
Fully irradiated & 0.950 & 0.253 & 1.945 & 0.134 & 0.528\\
\hspace{0.1cm} (grain size  $>75~\mu$m) & & & & &\\
\hline
\end{tabular}
\caption[Band Parameters for Eucrite Millbillillie]{Band Parameters for Eucrite Millbillillie}
\label{tab.laser}
\end{center}
\end{table}

\clearpage

\begin{table}
\begin{center}
\begin{tabular}{lccccc}
\hline 
\hline
Spectral Parameter & Cell 2 & Outside Cell 2 & $i<6^\circ$ & $i>6^\circ$ \\
\hline
BI Center		& 0.939	& 0.935	& 0.939	& 0.932 \\
BII Center		& 1.989	& 1.975	& 1.989	& 1.968 \\
BAR			& 2.182	& 2.346	& 2.170	& 2.369 \\
\hline
\end{tabular}
\caption[Median Asteroidal Band Parameters]{Median Asteroidal Band Parameters\\
This table lists the median spectral parameters for four different dynamical regions. See the text for definitions of these regions. The maximum statistical errors (as determined from the spectrum of asteroid 2000 OP49) for these band parameters are: $\sigma_{BI}=0.003$, $\sigma_{BII}=0.011$ and $\sigma_{BAR}=0.23$.
}
\label{tab.nesvorny}
\end{center}
\end{table}

\clearpage

\begin{center}
\begin{longtable}{lccccc}
\hline 
\hline
Sample Name & Type & Particle & BI Center & BII Center & BAR \\
 & & Size ($\mu$m) & ($\mu$m) & ($\mu$m) & \\
\hline
    Bialystok$^a$ & Howardite &  150 & 0.937 & 1.990 & 1.6414 \\
        Binda & Howardite &   25 & 0.928 & 1.948 & 1.6933 \\
       Bununu$^a$ & Howardite &   25 & 0.928 & 1.951 & 1.4740 \\
     EET83376 & Howardite &   25 & 0.934 & 1.981 & 1.7840 \\
     EET87503 & Howardite &   25 & 0.929 & 1.970 & 1.6192 \\
     EET87513 & Howardite &   25 & 0.932 & 1.976 & 1.4692 \\
    Frankfort$^a$ & Howardite &   25 & 0.927 & 1.945 & 1.7644 \\
     GRO95535 & Howardite &   25 & 0.929 & 1.963 & 1.6713 \\
     GRO95574 & Howardite &  125 & 0.929 & 1.953 & 1.7572 \\
      Kapoeta$^a$ & Howardite &   25 & 0.928 & 1.936 & 1.3722 \\
  Le Teilleul$^a$ & Howardite &   25 & 0.928 & 1.943 & 1.9352 \\
     Pavlovka$^a$ & Howardite & chip & 0.921 & 1.934 & 1.9778 \\
   Petersburg$^a$ & Howardite &   25 & 0.934 & 1.988 & 1.8304 \\
     QUE94200 & Howardite &   25 & 0.921 & 1.926 & 1.8620 \\
     QUE97001 & Howardite &  125 & 0.924 & 1.933 & 2.0792 \\
        Y7308 & Howardite &   25 & 0.927 & 1.942 & 1.6972 \\
      Y790727 & Howardite &   25 & 0.931 & 1.964 & 1.9103 \\
      Y791573 & Howardite &   25 & 0.926 & 1.948 & 1.8023 \\
       A87272 &   Eucrite &   25 & 0.941 & 2.029 & 1.2244 \\
      A881819 &   Eucrite &   25 & 0.931 & 1.976 & 1.5599 \\
     ALH78132 &   Eucrite &   45 & 0.933 & 1.967 & 2.1383 \\
     ALH85001 &   Eucrite &   25 & 0.925 & 1.952 & 1.5517 \\
    ALHA76005 &   Eucrite &   25 & 0.935 & 1.996 & 1.7261 \\
    ALHA81001 &   Eucrite &   45 & 0.937 & 2.024 & 2.7991 \\
    ALHA81011 &   Eucrite &  125 & 0.951 & 2.042 & 1.3355 \\
    ALHA85001 &   Eucrite & 1000 & 0.932 & 1.993 & 1.6795 \\
       Bereba$^a$ &   Eucrite &   25 & 0.941 & 2.023 & 1.7060 \\
     Bouvante &   Eucrite &   25 & 0.943 & 2.030 & 2.2411 \\
     BTN00300 &   Eucrite &   45 & 0.947 & 2.029 & 0.9420 \\
      Cachari &   Eucrite &   25 & 0.940 & 2.027 & 1.4802 \\
     EET83251 &   Eucrite & 1000 & 0.937 & 2.009 & 2.1906 \\
     EET87520 &   Eucrite &   45 & 0.951 & 2.042 & 1.1116 \\
     EET87542 &   Eucrite &   25 & 0.940 & 2.017 & 1.3458 \\
     EET90020 &   Eucrite &   25 & 0.941 & 2.012 & 0.9865 \\
     EET92003 &   Eucrite &  125 & 0.933 & 2.002 & 1.6588 \\
    EETA79005 &   Eucrite &   25 & 0.934 & 1.974 & 2.0033 \\
    EETA79006 &   Eucrite &  125 & 0.935 & 1.987 & 1.9342 \\
     EETA790B &   Eucrite & 1000 & 0.937 & 1.998 & 2.0071 \\
     GRO95533 &   Eucrite &   25 & 0.941 & 2.040 & 1.4581 \\
       Ibitra$^a$ &   Eucrite &   25 & 0.940 & 2.000 & 1.1341 \\
       Jonzac$^a$ &   Eucrite &   25 & 0.939 & 2.013 & 1.7952 \\
      Juvinas$^a$ &   Eucrite &   25 & 0.936 & 2.003 & 1.6049 \\
     LEW85303 &   Eucrite &   25 & 0.945 & 2.022 & 1.5695 \\
     LEW87004 &   Eucrite &   25 & 0.934 & 1.963 & 1.7839 \\
     MAC02522 &   Eucrite &   45 & 0.970 & 2.160 & 0.8309 \\
     MET01081 &   Eucrite &   45 & 0.939 & 2.002 & 1.2518 \\
Millbillillie$^a$ &   Eucrite &   25 & 0.938 & 2.019 & 1.3609 \\
  Moore County$^a$ &   Eucrite &   25 & 0.938 & 1.987 & 1.3885 \\
       NWA011 &   Eucrite &   25 & 0.950 & 2.029 & 0.9193 \\
 Padvarninkai$^a$ &   Eucrite &   25 & 0.939 & 2.012 & 1.4851 \\
    Pasamonte$^a$ &   Eucrite &   25 & 0.939 & 2.003 & 1.5453 \\
     PCA82501 &   Eucrite &  125 & 0.943 & 2.025 & 1.9056 \\
     PCA82502 &   Eucrite &   25 & 0.941 & 2.025 & 2.2161 \\
     PCA91006 &   Eucrite &  125 & 0.942 & 2.002 & 1.3592 \\
     PCA91007 &   Eucrite &  125 & 0.943 & 2.042 & 2.1384 \\
     PCA91078 &   Eucrite &   45 & 0.955 & 2.059 & 1.7840 \\
  Serra de Mage$^a$ &   Eucrite &   25 & 0.931 & 1.974 & 1.5117 \\
     Stannern$^a$ &   Eucrite &   25 & 0.938 & 2.017 & 1.9172 \\
       Y74450 &   Eucrite &   25 & 0.934 & 1.970 & 1.3623 \\
       Y75011 &   Eucrite & 1000 & 0.947 & 2.025 & 1.2378 \\
      Y791186 &   Eucrite & 1000 & 0.949 & 2.052 & 1.4823 \\
      Y792510 &   Eucrite &   25 & 0.942 & 2.032 & 1.3369 \\
      Y792769 &   Eucrite &   25 & 0.941 & 2.026 & 1.6585 \\
      Y793591 &   Eucrite &   25 & 0.939 & 2.011 & 1.8571 \\
       Y82082 &   Eucrite &   25 & 0.947 & 2.034 & 1.6305 \\
      Y980318 &   Eucrite &   75 & 0.940 & 2.016 & 1.4857 \\
    EETA79002 & Diogenite &   25 & 0.917 & 1.887 & 1.7581 \\
     Ellemeet$^a$ & Diogenite &   25 & 0.920 & 1.901 & 1.9083 \\
    Johnstown$^a$ & Diogenite &   25 & 0.914 & 1.874 & 1.5085 \\
     LAP91900 & Diogenite & 1000 & 0.925 & 1.920 & 1.9701 \\
         Roda$^a$ & Diogenite & chip & 0.923 & 1.904 & 1.9880 \\
       Shalka$^a$ & Diogenite & chip & 0.918 & 1.905 & 2.2039 \\
   Tatahouine$^a$ & Diogenite & chip & 0.920 & 1.916 & 1.8623 \\
       Y74013 & Diogenite &   25 & 0.920 & 1.919 & 1.7630 \\
       Y75032 & Diogenite &   25 & 0.926 & 1.924 & 1.7040 \\
\hline
\caption[]{Band Parameters of HED Meteorites\\
$^a$ Indicates a fall rather than a find.
}
\label{tab.ap1.HEDs}
\label{lasttable}
\end{longtable}
\end{center}

\clearpage


\clearpage

\begin{figure}
\begin{center}
Figure \ref{fig.obs} continued
\includegraphics[width=14cm]{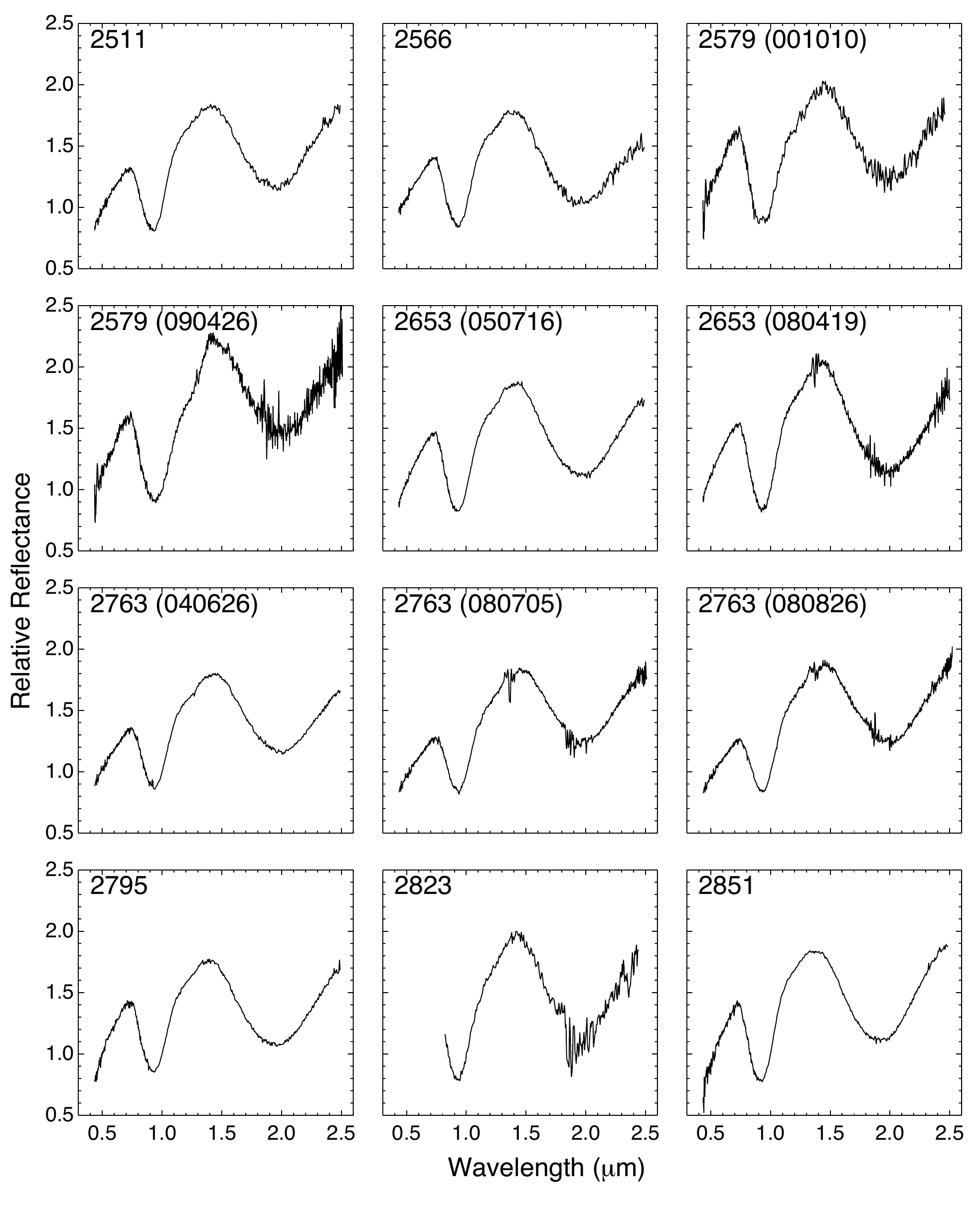}
\end{center}
\end{figure}
\begin{figure}
\begin{center}
Figure \ref{fig.obs} continued
\includegraphics[width=14cm]{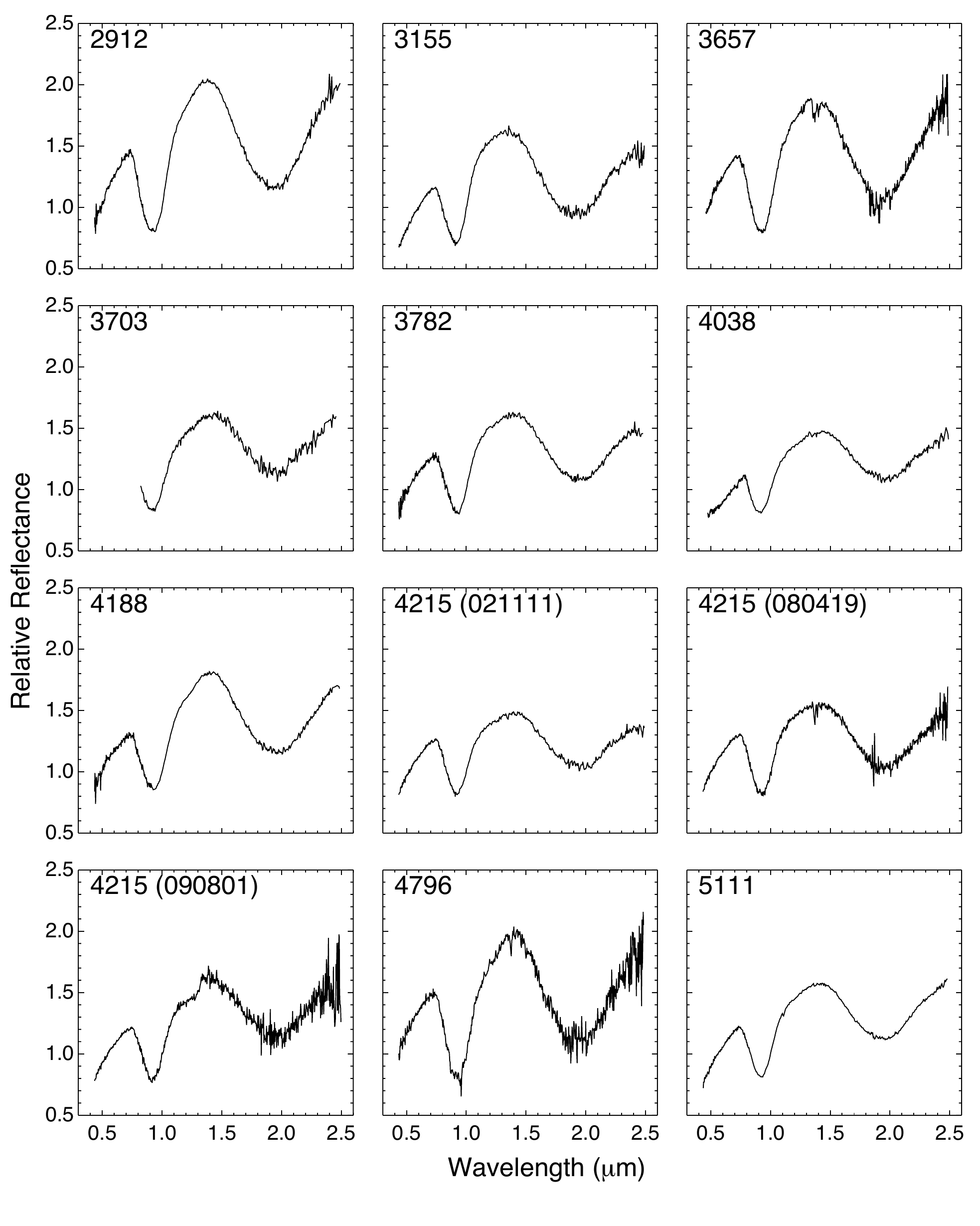}
\end{center}
\end{figure}
\begin{figure}
\begin{center}
Figure \ref{fig.obs} continued
\includegraphics[width=14cm]{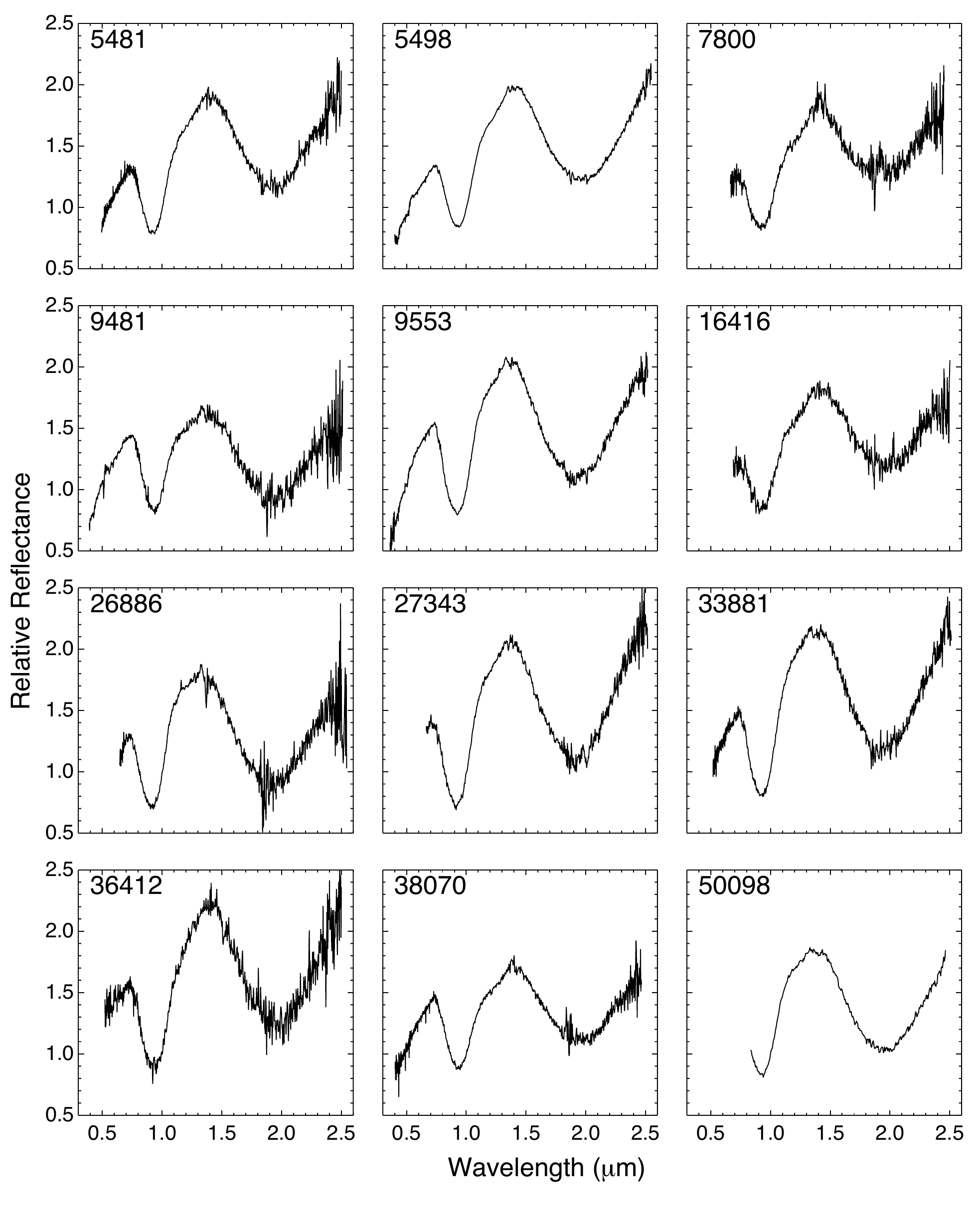}
\end{center}
\end{figure}
\begin{figure}
\begin{center}
Figure \ref{fig.obs} continued
\includegraphics[width=14cm]{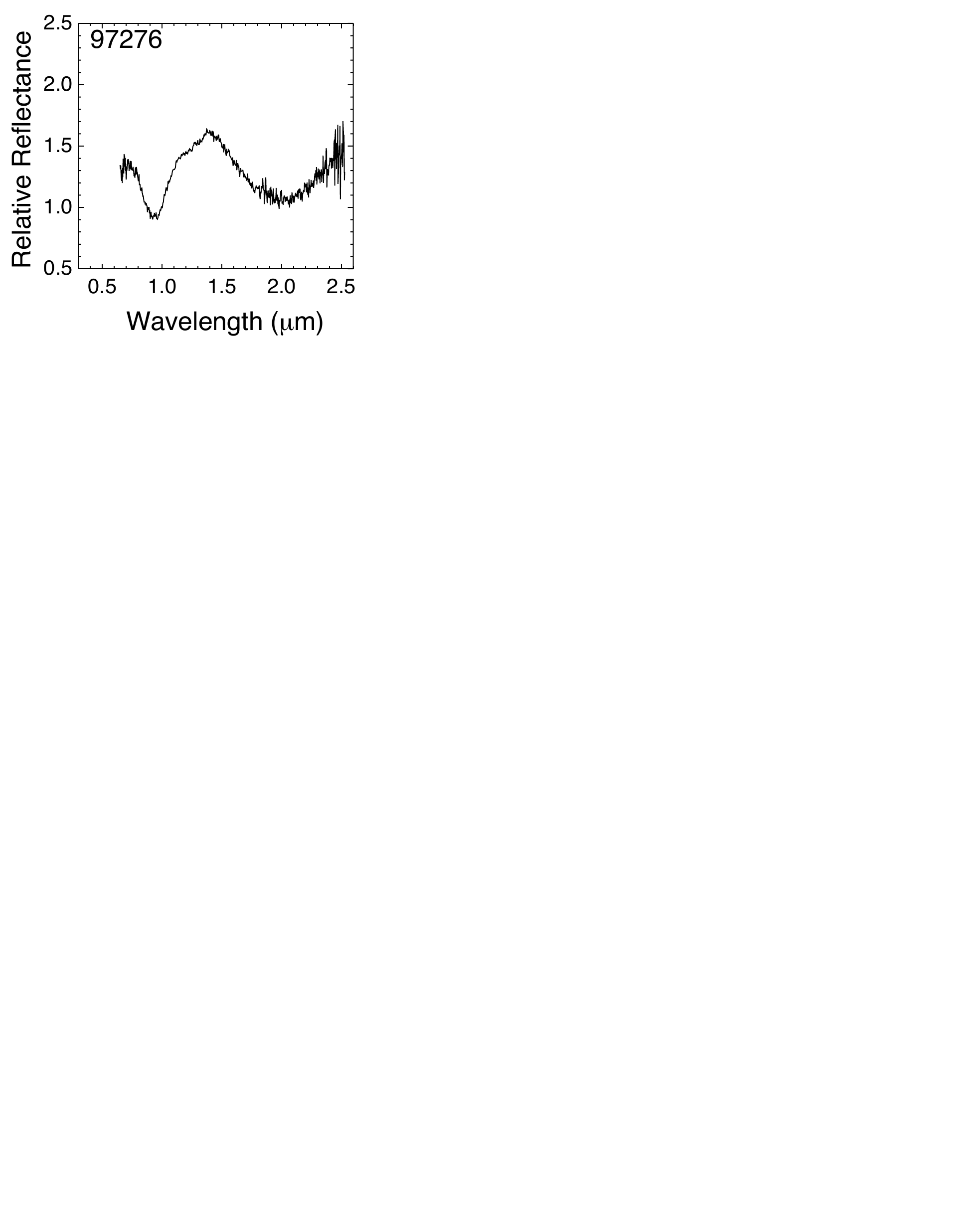}
\end{center}
\end{figure}

\end{document}